\documentclass[reprint,aps,pre,amsmath,amssymb,nopac]{revtex4-1}

\usepackage{amsfonts,amssymb,bm}
\usepackage{graphicx,epstopdf,epsfig,color}

\usepackage{dcolumn}
\usepackage{bm}
\newcommand{\bx}{{\bf x}}
\newcommand{\by}{{\bf y}}
\newcommand{\bz}{{\bf z}}
\newcommand{\bv}{{\bf v}}
\newcommand{\br}{{\bf r}}
\newcommand{\bn}{{\bf n}}

\newcommand{\bq}{{\bf q}}

\newcommand{\rd}{{\rm d}}
\newcommand{\rh}{{\rm h}}

\begin{document}
\title{Network and geometric characterization of three-dimensional fluid transport between two layers
}
\author{Rebeca de la Fuente$^1$, G\'{a}bor Dr\'{o}tos$^{1,2}$,
Emilio Hern\'andez-Garc\'{\i}a$^1$, Crist\'obal L\'opez$^1$}

\affiliation{$^1$IFISC (CSIC-UIB). Instituto de F\'{\i}sica
Interdisciplinar y Sistemas Complejos,
E-07122 Palma de Mallorca, Spain \\
$^2$ MTA--ELTE Theoretical Physics Research Group, Budapest,
Hungary}
\date{\today}

\begin{abstract}

We consider transport in a fluid flow of
arbitrary complexity but with a dominant flow direction. One of
the situations in which this occurs is when describing by an
effective flow the dynamics of sufficiently small particles
immersed in a turbulent fluid and vertically sinking because of
their weight. We develop a formalism characterizing the
dynamics of particles released from one layer of fluid and
arriving in a second one after traveling along the dominant
direction. The main ingredient in our study is the definition
of a two-layer map that describes the Lagrangian transport
between both layers. We combine geometric approaches and
probabilistic network descriptions to analyze the two-layer
map. From the geometric point of view, we express the
properties of lines, surfaces and densities transported by the
flow in terms of singular values related to Lyapunov exponents,
and define a new quantifier, the Finite Depth Lyapunov
Exponent. Within the network approach, degrees and an entropy
are introduced to characterize transport. We also provide
relationships between both methodologies. The formalism is
illustrated with numerical results for a modification of the
ABC flow, a model commonly studied to characterize
three-dimensional chaotic advection.

\end{abstract}

\maketitle

\section{Introduction}

The study of transport phenomena is at the core of fluid
mechanics. The Lagrangian approach to fluid transport has
received powerful insights from its relationship to chaos and
dynamical systems \cite{ottino1989kinematics,doi:10.1146/annurev.fluid.37.061903.175815,Shadden2005}, and more
recently from set-oriented methodologies which can be recast
into the language of graph or network theory
\cite{bolltbook2013,SerGiacomi2015,doi:10.1063/1.4892530,PhysRevLett.98.224503}.

In most of the previous applications of these developments to
geophysical contexts, consideration has been restricted to
horizontal transport, as this is the dominant mode of motion at
large scales in oceans and in the atmosphere. Some works,
however, have addressed the full three-dimensional dynamics
\cite{Froyland2015,bettencourt2015boundaries,https://doi.org/10.1002/2017JC012700}.
 Less attention has
been given to the application or adaptation of the approaches
mentioned above to the peculiarities of transport in the
vertical direction, which is singled out by the gravitational
force.

As the main motivation for the present work, many relevant
biogeochemical phenomena involve the vertical transport of
particles in the ocean. Two paradigmatic examples are the
sinking of biogenic particles \cite{Siegel1997,Monroy2017},
like phytoplankton cells and marine snow, which play a
fundamental role in the biological carbon pump \cite{Sabine367,DELAROCHA2007639}, and the
sedimentation dynamics of microplastics, which are becoming a
key environmental problem \cite{choy2019vertical,doi:10.1021/acs.est.0c01984,delaFuente2021}.
Despite the numerous studies with different experimental and
theoretical methodologies many questions remain open, in
particular those concerning the final fate of the particles
from a known release surface area (i.e. the connection paths
between surface and deep ocean), the amount and time they are
suspended in the water column, and the spatial distribution
both over the water column and the seafloor. Beyond the ocean
context, vertical transport is also relevant in many other
situations such as engineering processes \cite{10.1115/1.1537258} or rain precipitation \cite{falkovich2002acceleration}.
%

The objective of this paper is to extend and adapt the powerful
previously commented Lagrangian methodologies
to dynamics for which there is a strong
anisotropy in the particle motion, leading to a clear transport
direction. This is the case when considering sinking particles
in fluid flows. We will concentrate on characterizing
transport between two layers: in the case of particles
sedimenting under gravity, particles released from an upper
layer are driven by the flow and reach and accumulate in a
lower layer. We expect our formalism would be useful also under
transport anisotropies produced by forces other than gravity.
The main object we will define is a two-layer map that connects
the initial conditions of particles released from one of the
layers to their final positions in the other one, after being
transported by the flow. We extract information from this map
with the two complementary approaches mentioned above: on the
one hand we use dynamical systems tools to describe the
geometry of the evolution of sheets of particles released from
the initial layer. In this way we formalize previous results
obtained in this context
\cite{Drotos2019,Monroy2019,Sozza2020} and
extend them
 by the introduction of a new quantifier related to
Lyapunov exponents: the Finite Depth Lyapunov Exponent. On the
other hand, connectivity properties between the layers are
studied with network theory or probabilistic techniques.
Relationships between both approaches are obtained,
and the whole formalism is illustrated with a modification of the ABC flow. This flow model
is frequently used as a simple example of three-dimensional chaotic advection, to which we
add an additional constant velocity in the vertical direction to model sinking.

The outline of the paper is as follows. In Section
\ref{Sec:transport} we introduce the basic Lagrangian
description for transport of particles between two layers. In
\ref{Sec:Falling} we study the geometry and dynamics of a
falling layer of particles, introducing the new type of
Lyapunov exponent. In Section \ref{Sec:Net} we introduce the
network methods to characterize connectivity, and in Section
\ref{Sec:relation} we show the connection between the previous
two descriptions. In Section \ref{Sec:num} we present the
numerical results obtained for the modified ABC flow model.
Section \ref{Sec:con} presents our conclusions. An Appendix
contains additional technical details.

\section{Characterization of transport between two layers}
\label{Sec:transport}

Given a fluid flow characterized by a velocity field
$\bv(\br,t)$, the Lagrangian description of transport considers
the equations of motion for the position of fluid elements,
which evolve according to
\begin{equation}
\frac{\rd{\br}(t)}{\rd t}=\bv (\br (t),t) \ .
\label{eq:lag}
\end{equation}
This equation defines the flow map
$\phi_{t_{0}}^{\tau}(\br_{0})$, such that integrating Eq.
(\ref{eq:lag}) for a given initial condition \textbf{$\br_{0}$}
at $t_{0}$ gives the final position of the particle at
time $t_{0}+\tau$:
\begin{equation}
\phi_{t_{0}}^{\tau}(\br_{0})=\br(t_{0}+\tau).
\label{eq:lagmap}
\end{equation}
In the rest of the paper, we will restrict to the situation in
which $\bv (\br (t),t)$ is a three-dimensional velocity field,
and trajectories $\br(t)$ move in regions of $\mathbb{R}^3$.

Description (\ref{eq:lag}) is not only pertinent for the motion
of fluid elements. Particles of other substances immersed in a
fluid also satisfy a first-order equation like
(\ref{eq:lag}), provided they are sufficiently small for their
inertia to be neglected. For example, in a variety of realistic
situations in the ocean, the equation of motion for the
position of many types of particles of biological origin or of
microplastics is ruled by Eq. (\ref{eq:lag}),
in which the velocity field is replaced
by the actual velocity of the fluid flow with an added
constant vertical component
related to
the sinking of the particle under gravity because of its
weight \cite{Siegel1997,Monroy2017,delaFuente2021}.
In this
paper we will refer to
 the motion of `particles' without specifying if
 they are particles of fluid or particles submerged in a fluid.
 In both cases the dynamics is provided by an equation
of the type (\ref{eq:lag}), and thus Eq. (\ref{eq:lagmap}) applies.

An object that plays an important role in the analysis of the
map in (\ref{eq:lagmap}) is its Jacobian matrix (a $3\times 3$
matrix), defined by
\begin{equation}
\bm{J} = \nabla \phi_{t_{0}}^{\tau}(\br_{0}).
\label{eq:jac_t}
\end{equation}
Given an  infinitesimal separation between two initial
conditions $\rd\br_0$, $\bm{J}$ gives the evolution in time of this
separation: $\rd\br(t_0+\tau)=\bm{J}\cdot \rd\br_0$. The singular values
$\{S_\alpha\}_{\alpha=1,2,3}$ of $\bm{J}$ (i.e. the square roots of
the eigenvalues of the Cauchy-Green tensor $C = \bm{J}^T \bm{J}$) give
the stretching factors experienced by infinitesimal material
line elements oriented along the eigendirections and started
around $\br_0$ while integrated from $t_{0}$ to $t_0+\tau$. The
standard finite-time Lyapunov exponents (FTLE,
$\{\lambda_\alpha\}_{\alpha=1,2,3}$) are obtained from these
singular values as $\lambda_\alpha=|\tau|^{-1}\ln S_\alpha$
\cite{Shadden2005}.

In this paper we are interested in anisotropic situations in
which a direction of flow is distinguished from the others.
Specifically, instead of the fully three-dimensional
motion described by $\phi_{t_{0}}^{\tau}$, we are interested in
the dynamics of particles traveling between a pair of
two-dimensional layers. The main example is the case of
particles released from an upper horizontal layer, falling by
gravity across a moving flow, and being collected on a second
lower horizontal layer.
Other sources of anisotropy can play
the role of gravity, but in this paper we use the terminology appropriate to the sedimentation
by gravity example, so that both layers will be considered to be horizontal. The first layer
will be called the \textit{upper} or \textit{release} layer, whereas the second one will
 be called the \textit{lower} or the \textit{collecting} layer.
We distinguish
the \textit{vertical} coordinate $z$ from the
\textit{horizontal} ones that form the horizontal vector $\bx$,
so that $\br=(x,y,x) \equiv (\bx,z)$. Particles are initially
released (at $t_0$) from the horizontal layer $\mathcal{M}$
characterized by `height' $z_0$: $\mathcal{M}\equiv
\{\br=(\bx_0,z_{0}), z_0\textrm{ fixed}\}$, and we want to
track the horizontal position $\bx$ at which the particle
started at $\br_0$ first reaches the second horizontal layer
characterized by `depth' $z$. As we stop the dynamics after
this first arrival, we can say that particles `accumulate' at
the second layer. This procedure defines a new flow map which
we call the \textit{two-layer map}: $\bx=\phi_{z_0}^z (\bx_0)$.
We do not explicitly specify the initial time $t_0$ but for
time-dependent velocity fields there will be a
dependence on it.

Given a region $D\in\mathcal{M}$ of the upper layer, we call
its image $\phi_{z_0}^z (D)$ onto the lower one its
\textit{footprint}. It is the region of the collecting layer
where particles from $D$ will become accumulated.

Particles released at the same time do not necessarily arrive
at the same time at the final layer. Let $\omega(\bx_0)$ be the
time that a particle started at $t_0$ from $(\bx_0,z_0)$ takes
to reach the second layer at $z$ for the first time. Thus the
time of arrival is $t_z=t_0+\omega$. Although not explicitly
written, $\omega$ and $t_z$ depend on $t_0$, $z_0$ and $z$, in
addition to $\bx_0$. In terms of $\omega$, the relationship
between the coordinates of the two flow maps introduced so far
is:
\begin{eqnarray}
\phi_{z_0}^z (\bx_0)= \bx (t_0 + \omega(\bx_0) )
   &=&
\phi_{t_0}^{\omega(\bx_0)}\left(\br=(\bx_0,z_0)\right)\left.\right|_\rh
\nonumber, \\
z &=& \phi_{t_0}^{\omega(\bx_0)}\left(\br=(\bx_0,z_0)\right)
\left.\right|_z,
\label{eq:rel}
\end{eqnarray}
where the subindices $\rh$ and $z$ indicate that the horizontal
and vertical coordinates of $\phi_{t_0}^\omega$, respectively,
should be taken.

In general
$\phi_{z_0}^z$ can always be computed by solving Eq.
(\ref{eq:lag}) from initial conditions on $\mathcal{M}$, and
checking when the trajectory crosses the second layer at $z$,
as Eq. (\ref{eq:rel}) indicates. In this paper we will use this
last method.

The Jacobian associated with the two-layer map is
\begin{equation}
\bar{\bm{J}}_{\mathcal{M}} = \nabla \phi_{z_0}^z (\bx_0) \ .
\label{eq:M}
\end{equation}
Note that the gradient acts on the two-dimensional initial
position $\bx_0$, so that $\bar{\bm{J}}_{\mathcal{M}}$ is a $2\times
2$ matrix. The subindex ${\mathcal{M}}$ is a reminder of the
fact that $\bar{\bm{J}}_{\mathcal{M}}$ is defined on each point
$\bx_0$ of the upper layer $\mathcal{M}$.

The singular values of this new Jacobian matrix are the square
roots of the eigenvalues of the associated $2\times 2$
Cauchy-Green tensor:
\begin{equation}
\bar C_{\mathcal{M}} = (\nabla \phi_{z_{0}}^{z}(\bx_0))^{T} \cdot \nabla \phi_{z_{0}}^{z}(\bx_0)  ,
\label{eq:barCG}
\end{equation}
which will be used later on.

We next develop the two complementary approaches we propose to
study transport between two layers: the geometric and the
network approaches.

\section{Geometric characterization of a falling layer}
\label{Sec:Falling}

First we introduce a geometric characterization of the
deformation of the falling layer of released particles with
tools from dynamical systems. This approach can be called both
\textit{geometric} or \textit{dynamical}.

In the same way that the three-dimensional Jacobian matrix $\bm{J}$
maps infinitesimal vector particle separations from time $t_0$
to time $t$ ($\rd\br(t)=\bm{J}\cdot \rd\br_0$), ${\bm{\bar{J}}}_\mathcal{M}$
takes initial infinitesimal separations $\rd\bx_0$ on the
horizontal release layer and gives its footprint $\rd\bx_z$ on
the collecting layer: $\rd\bx_z={\bm{\bar{J}}}_\mathcal{M}\cdot \rd\bx_0$.
The singular values $\bar\Lambda_1$ and $\bar\Lambda_2$ of
${\bm{\bar{J}}}_\mathcal{M}$ give the stretching factors experienced by the footprint of
line elements initially oriented along the eigendirections of
$\bar C_\mathcal{M}$. In analogy with the definition of FTLEs,
we can define \emph{Finite-Depth Lyapunov Exponents} (FDLEs,
$\bar\lambda_\alpha(\bx_0)$) as the logarithmic rate of
stretching along the eigendirections:
\begin{equation}
\bar\lambda_\alpha(\bx_0) = \frac{1}{|z-z_0|}\log \bar \Lambda_\alpha \ , \alpha=1,2 .
\label{eq:FDLE}
\end{equation}
$\bar\lambda_\alpha$ is naturally expressed as a function of
$\bx_0$. But in fact it is a property of the trajectory joining
$\bx_0$ and $\bx=\phi_{z_0}^z(\bx_0)$, so that it (and also
$\bar\Lambda_\alpha$) can be thought and displayed as a
function of the coordinates on the collecting layer, $\bx$.
Although not explicitly indicated, $\bar\lambda_\alpha$ (and
$\bar\Lambda_\alpha$) is a function of $t_0$, $z_0$ and
$z$. Values $\bar\lambda_\alpha>0$ ($\bar\Lambda_\alpha>1$)
indicate growth of lengths initially oriented
along the corresponding eigendirection, whereas
$\bar\lambda_\alpha<0$ ($\bar\Lambda_\alpha<1$) indicate length
contraction. If $\bar\lambda_1>\bar\lambda_2$, for sufficiently
large differences of depth $|z-z_0|$ we would have $|\rd\bx_z|
\approx e^{|z-z_0|\bar\lambda_1} |\rd\bx_0^{(1)}|$, where
$\rd\bx_0^{(1)}$ is the projection of the initial particle
separation $\rd\bx_0$ onto the singular vector of singular
value $\bar\Lambda_1$.

At difference with the FTLE, the FDLE has dimensions of inverse of
length, not of time. But this is not the most important
difference between the two quantities (in fact an alternative
definition could be to replace $|z-z_0|$ by $\omega$ in
(\ref{eq:FDLE})). The main difference is that the FTLE
quantifies the stretching of initial vectors as they are
transported by the flow in three-dimensional space, whereas the
FDLE also includes the projection effect experienced by these
vectors when arriving at the collecting layer: the footprint of such a vector is
the projection onto the horizontal layer of that vector arriving
there, taken along its direction of motion. Further details of
this projection process are given in the Appendix, and are also
illustrated in Fig. \ref{fig:ProjectionPlot}. Note also that
the FDLE is not a form of a finite-size Lyapunov exponent
\cite{Aurell_1997,Bettencourt_2013,Cencini_2013}, since for this last quantity initial
separations are integrated until reaching a specified
separation value, whereas in the FDLE integration proceeds
until reaching a particular depth level $z$.

Next, we consider the effect of the flow on
surface elements initially located in the release layer.
This was already considered in
\cite{Monroy2017,Drotos2019,Monroy2019,Sozza2020} in the
context of sedimenting particles in fluid flows.

\begin{figure}[h]
	\includegraphics[width=.8\columnwidth]{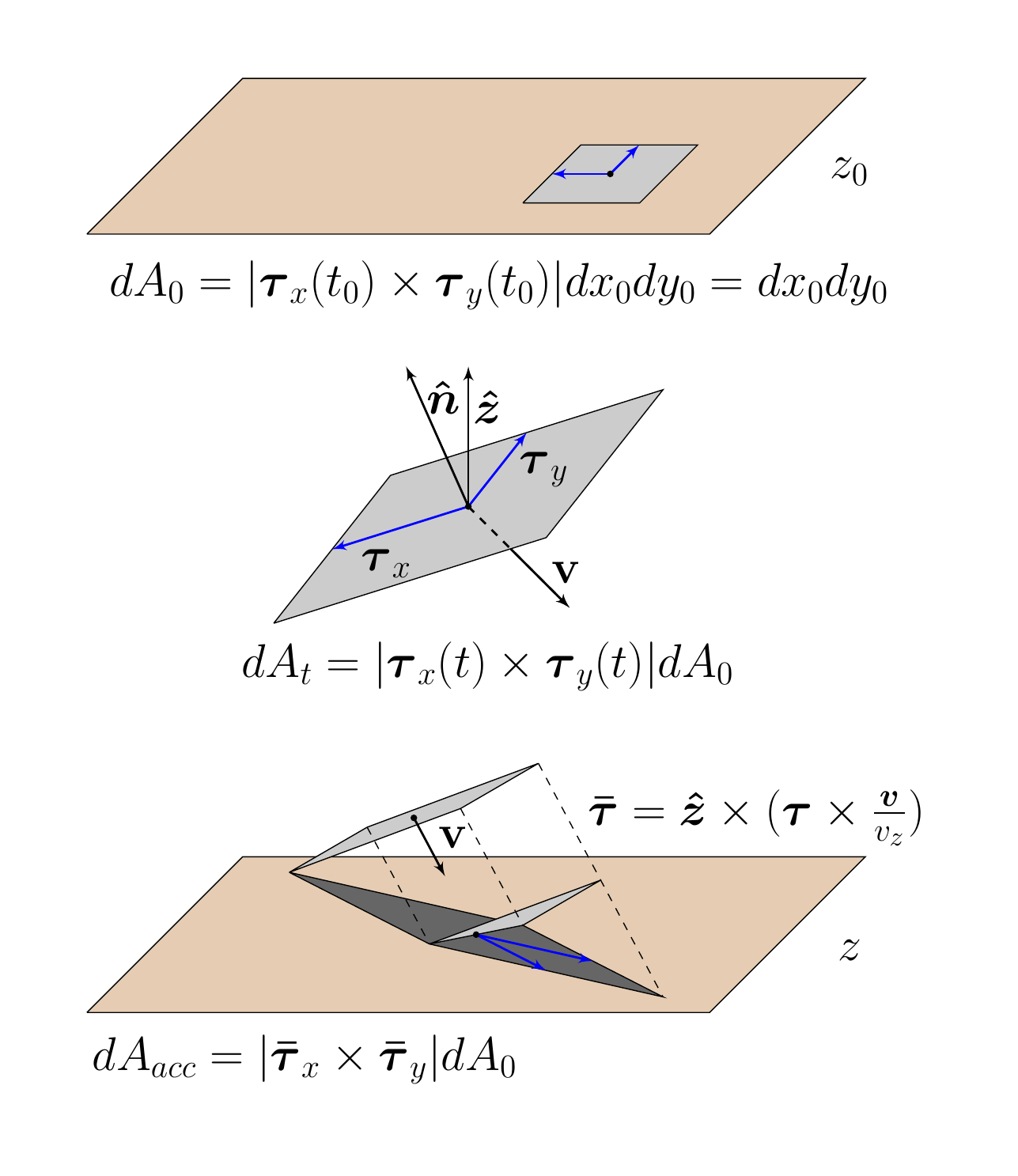}
	\caption{Illustration of the dynamics of
a rectangular surface element, lying on the upper layer at the release
time $t_0$, and with area $\rd A_0$, until leaving a footprint of area
$\rd A_\mathrm{acc}$ on the lower layer when arriving there.
See main text and Appendix for details. }
	\label{fig:ProjectionPlot}
\end{figure}

Let us consider an infinitesimal material surface of area
$\rd A_0$ started at the release layer at $z_0$, which at any time
is transformed into a surface of area $\rd A_t$, and which finally
reaches the collecting layer at $z$ leaving a footprint area
$\rd A_\mathrm{acc}$ (see Fig. \ref{fig:ProjectionPlot}).
If we take the
initial surface element to be a rectangle of sides given by the
vectors $\hat \bx \rd x_0$ and $\hat \by \rd y_0$ ($\hat \bx$ and
$\hat \by$ are unit vectors in the $x$ and $y$ directions; the
area of the rectangle is $\rd A_0=\rd x_0 \rd y_0$), and noting that the
cross product of vectors gives the area of the parallelogram
subtended by them, we obtain
\begin{equation}
\rd A_\mathrm{acc}= | {\bm{\bar{\tau}_x}}\times {\bm{\bar{\tau}_y}} | \rd A_0,
\label{eq:acumulada}
\end{equation}
where ${\bm{\bar{\tau}_x}}= \frac{\partial
\phi_{z_0}^z(\bx_0)}{\partial x_0} $ and
${\bm{\bar{\tau}_y}}=\frac{\partial
\phi_{z_0}^z(\bx_0)}{\partial y_0}$ are two-dimensional vectors
on the final layer such that ${\bm{\bar{\tau}_x}}\rd x_0$ and
${\bm{\bar{\tau}_y}}\rd y_0$ give the footprint of the initial
vectors $\hat \bx \rd x_0$ and $\hat \by \rd y_0$.

Simple algebra relates the cross product in
(\ref{eq:acumulada}) to the matrix $\bar C_\mathcal{M}$ and
the singular values $\bar\Lambda_\alpha$:
\begin{equation}
|{\bm{\bar{\tau}}_x} \times  {\bm{\bar{\tau}}_y} |=
\sqrt{\det \bar C_\mathcal{M}}=
\bar \Lambda_1 \bar \Lambda_2 \equiv F^{-1}  \ ,
\label{eq:bartau_barlambda}
\end{equation}
where we have defined the quantity $F$ which we call the
\emph{density factor}. It is a function of the trajectory that
starts at $\bx_0$ and arrives at $\bx=\phi_{z_0}^z(\bx_0)$, so
that, with some abuse of language, it can be considered either
as a function of the initial or of the final location:
$F=F(\bx_0)$ or $F=F(\bx)$. The name density factor comes from
the consideration of the ratio between the density of particles
in a release surface element, $\sigma(\bx_0)$, and in its image
in the collecting layer $\sigma(\bx=\phi_{z_0}^z(\bx_0))$. In
the situation in which both surface elements contain the same particles, this ratio is the inverse of the ratio of
areas, and thus equal to $F$:
\begin{equation}
\frac{\sigma(\bx)}{\sigma(\bx_0)}=\frac{\rd A_0}{\rd A_\mathrm{acc}} = F.
\label{eq:df}
\end{equation}
The surface elements $\rd A_0$ and $\rd A_\mathrm{acc}$ will contain the
same particles if a single surface element from the release
layer reaches $\rd A_\mathrm{acc}$. For time-dependent velocity fields,
folding of the falling layer can occur, and in this case the
complete density ratio should be computed as the sum of all
contributions of the type (\ref{eq:df}) from the initial
release areas $\rd A_0$ that reach the same $\rd A_\mathrm{acc}$ at
different times \cite{Drotos2019,Monroy2019,Sozza2020}.

A convenient way to write $F=\rd A_0/\rd A_\mathrm{acc}$ is to split it into
two contributions \cite{Drotos2019,Monroy2019,Sozza2020}
 (see Fig. \ref{fig:ProjectionPlot}): the
evolution of the surface element under the time map
$\phi_{t_0}^\omega$ until when its area gets stretched to
$\rd A_{t_z}$ (recall that $t_z=t_0+\omega$ is the time at which
the infinitesimal surface touches the $z$ layer),
and the projection of this surface element onto the horizontal collection
layer along the direction of motion. The combination of both processes leaves a footprint of
area $\rd A_\mathrm{acc}$ on the bottom layer, completing the action of $\phi_{z_0}^{z}$:
\begin{equation}
F= \frac{\rd A_0}{\rd A_{t_z}} \frac{\rd A_{t_z}}{\rd A_\mathrm{acc}} = S~P \ .
\label{eq:eqF}
\end{equation}
The stretching and projection factors, $S$ and $P$, can be
calculated as \cite{Drotos2019,Monroy2019,Sozza2020}:
\begin{eqnarray}
S= \frac{\rd A_0}{\rd A_{t_z}} &=& \left|\bm{\tau}_x (t_z) \times \bm{\tau}_y (t_z) \right|^{-1},
\label{eq:SyP1}\\
P = \frac{\rd A_{t_z}}{\rd A_\mathrm{acc}} &=& \left|\frac{v_z}{\hat\bn \cdot \bv}\right|,
\label{eq:SyP2}
\end{eqnarray}
where $\hat\bn (t)$ is a unit vector normal to the falling surface
element at time $t$, and the vectors $\bm{\tau}_x (t)$ and
$\bm{\tau}_y (t)$ are tangent to the sinking surface $\rd A_t$ at
time $t$, given by
$\bm{\tau}_x=\frac{\partial{\phi_{t_0}^\tau}(\br_0)}{{\partial x_0
}}$ and
$\bm{\tau}_y=\frac{\partial{\phi_{t_0}^\tau(\br_0)}}{{\partial y_0
}}$. The expression for $S$ is obtained simply by recognizing
that $\bm{\tau}_x(t) \rd x_0$ and $\bm{\tau}_y(t) \rd y_0$ are the
images under time evolution of the vectors $\bm{\hat x}\rd x_0$
and $\bm{\hat y}\rd y_0$, respectively, that make the initial
surface, and thus the area at any time $t$ is
$\rd A_t=|\bm{\tau}_x (t) \times \bm{\tau}_y (t)| \rd x_0 \rd y_0$. A derivation
of the expression for $P$ is given in the Appendix, where
further details on the projection process is given.
As with $\lambda_\alpha$, expression (\ref{eq:SyP1}) is a
property of the trajectory joining $\bx_0$ and the
corresponding $\bx$ in the collecting layer, so that $S$ can be
considered as a function of any of these two locations. Eq.
(\ref{eq:SyP2}) involves velocities and the normal to the surface
element at the collecting layer, so that it is more natural to
consider $P=P(\bx)$, although for invertible $\phi_{z_0}^z$ the
values of $P$ can also be mapped back to the release layer and
displayed there.

The density factor $F$ can also be expressed in terms of
singular values of a different Jacobian matrix. We begin with expressing the stretching factor $S$. First note that
the Jacobian matrix in (\ref{eq:jac_t}) has as columns the two
vectors $\bm{\tau}_x (t)$, $\bm{\tau}_y (t)$, and the
additional one $\bm{\tau}_z (t)
=\frac{\partial{\phi_{t_0}^\tau}(\br_0)}{{\partial z_0 }}$. Let
$\bm{J}_\mathcal{M}$ be the $3\times 2$ matrix having as columns
just the three-dimensional vectors $\bm{\tau}_x (t)$,
$\bm{\tau}_y (t)$. The subindex $\mathcal{M}$ indicates that it
involves derivatives only along the horizontal release layer
$\mathcal{M}$. The singular values of $\bm{J}_\mathcal{M}$,
$\Lambda_1$ and $\Lambda_2$, are the square roots of the
eigenvalues of the $2\times 2$ matrix $C_\mathcal{M} =
\bm{J}_\mathcal{M}^T \bm{J}_\mathcal{M}$. Simple algebra demonstrates
that
\begin{equation}
S^{-1}=
|{\bm{{\tau}}_x}(t) \times  {\bm{{\tau}}_y}(t) |=
\sqrt{\det C_\mathcal{M}}=
\Lambda_1 \Lambda_2 \ .
\label{eq:tau_lambda}
\end{equation}
We stress that the quantities $\Lambda_\alpha$ are in general
different from the singular values $S_\alpha$ of the $3\times
3$ matrix $\bm{J}$ in Eq. (\ref{eq:jac_t}), giving the Lyapunov
exponents as $\lambda_\alpha=|\omega|^{-1}\log S_\alpha$.
$\Lambda_\alpha$ characterizes stretching only of infinitesimal
initial vectors lying on the horizontal initial layer. But, in
the limit of large $t_z$ or $|z-z_0|$, vectors of arbitrary
initial orientation are expected to approach the directions
that stretch faster under the action of $\bm{J}$, so that we expect
that in this limit $\Lambda_\alpha$ will approach $S_\alpha$,
for $\alpha=1,2$. More in general, since $\bm{J}_\mathcal{M}$ is the
matrix $\bm{J}$ with a column deleted, inequalities for singular
values of submatrices \cite{MatrixAnalysis} lead to
$S_{\alpha+1}\leq \Lambda_\alpha \leq S_\alpha$, with
$\alpha=1,2$.

Comparison of Eqs.~(\ref{eq:bartau_barlambda}), (\ref{eq:eqF})
and (\ref{eq:tau_lambda}) gives the following relationship
between the descriptions based on the singular values of $\bar{J}_\mathcal{M}$ and $J_\mathcal{M}$:
\begin{equation}
\bar \Lambda_1 \bar \Lambda_2
= P^{-1} \Lambda_1 \Lambda_2 = F^{-1},
\label{eq:releigen}
\end{equation}
which also shows the two different ways to compute the density factor $F$.

\section{The network approach}
\label{Sec:Net}

We now describe a characterization of fluid transport between
layers by tools from network or graph theory. This type of
approach can also be called \textit{probabilistic}, or
\textit{set-oriented}. Our goal is to generalize studies such
as \cite{Froyland2003,SerGiacomi2015} by
considering a bipartite network which is the natural framework
to study two-layer transport. For this we construct the
discrete version of the Perron-Frobenius operator describing
the transport matrix between the two layers.

\subsection{Coarse-graining of the flow and transport matrix}
\begin{figure}[h]
 	\includegraphics[width=0.8\columnwidth]{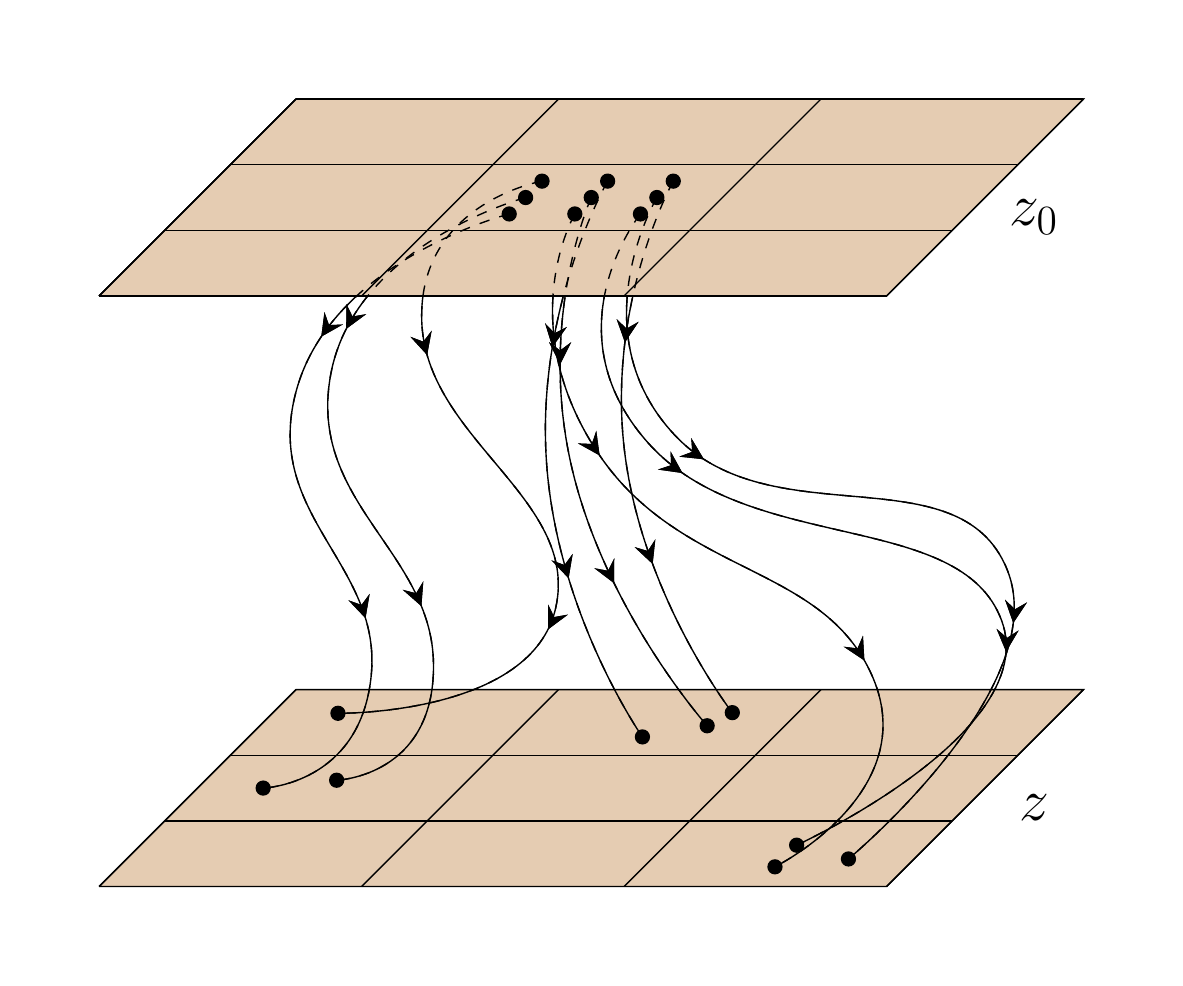}
	\caption{Sketch of the bipartite network construction. Particles
	travel from the upper layer to the bottom one. Nodes are the boxes
$A_i$, $i=1,...,M_0$ on which the upper layer is partitioned, and $B_j$, $j=1,...,M_z$,
partitioning the lower one. Two nodes are linked if some trajectory joins them.}
	\label{fig:twolayers}
\end{figure}

The upper layer is partitioned with a set of boxes $\{ A_{i}
\}_{i=1,\ldots,M_0}$, and the lower layer with boxes $\{ B_{j}
\}_{j=1,\ldots,M_z}$ (see Fig. \ref{fig:twolayers}). Each of
these boxes is interpreted as a node in a bipartite network.
Links between the upper and the lower layer are established by
the action of the two-layer map. These links are directed and
weighted, with weights between $A_{i}$ in the upper layer and
$B_{j}$ in the lower one given by the proportion of area of
$A_{i}$ which is mapped onto $B_{j}$, which defines a transport
matrix:

\begin{equation}
\textbf{P}(z_{0},z)_{ij}=\frac{\mu \left( A_{i} \bigcap
\left(\phi^{z}_{z_0}\right)^{-1}(B_{j})\right)}{\mu(A_{i})}.
\label{eq:Pij}
\end{equation}

$\mu(S)$ is the measure of set $S$ (a part of the release
layer) here taken to be its area. The map
$\left(\phi^{z}_{z_0}\right)^{-1}$ is the inverse of
$\phi^{z}_{z_0}$, i.e. it takes points from the lower layer
that at some moment were reached by the released particles and
maps them back into the position they had at $t_0$ in the upper
layer. Note that for time-dependent velocity fields this
inverse map can be multivalued, as several initial conditions
$\bx_0$ can reach the same point in the lower layer, provided
they do so at different times. In this case, all preimages of
box $B_j$ should be considered in Eq. (\ref{eq:Pij}). From a
practical point of view, one computes the matrix elements
$\textbf{P}(z_{0},z)_{ij}$ by releasing a large number $N_i$ of
trajectories from box $A_i$ at $t_0$, and counts how many of
them, $N_{ij}$, reach the collecting layer for the first time
at box $B_j$. The ratio $N_{ij}/N_i$ estimates the value of
$\textbf{P}(z_{0},z)_{ij}$ for $N_i$ large enough.

We note that the transport matrix $\textbf{P}(z_{0},z)$ is
different from the one used in previous works in two aspects:
first, it represents connections between two distinct regions:
the release and the collecting layer, whereas the transport
matrix used for example in
\cite{Froyland2003,SerGiacomi2015} quantifies the
transport between boxes embedded in the same fluid region. This
bipartite character of our transport matrix is shared by other
operators in the literature, for example
\cite{Froyland2015},
but then the second
difference is that in those cases transport is computed during
a fixed amount of time, whereas in our case what is fixed is
the distance between the two layers, with possibly different
times of transport between them for different particles.

Eq. (\ref{eq:Pij}) immediately leads to a probabilistic
interpretation: $\textbf{P}(z_{0},z)_{ij} \geq 0$ is the
probability that a particle started at $t_0$ in a
uniformly random position in box $A_i$ of the release layer
reaches the collecting layer for the first time on box $B_j$.
If all the particles released from $\mathcal{M}$ reach the
collecting layer, then $\textbf{P}(z_{0},z)_{ij}$ is row
stochastic:
\begin{equation}
\sum_{j=1}^{M_z} \textbf{P}(z_{0},z)_{ij} = 1.
\end{equation}
If some particles never reach the collecting layer, then
we can have $\sum_{j=1}^M \textbf{P}(z_{0},z)_{i,j} < 1$, being
this the probability of reaching the lower layer if starting
from a random position in the release one. As when dealing with
open flows \cite{ser2017lagrangian} one can
consider the transport matrix that takes into account only the particles that
do reach the second layer. The so-called
\textit{out-strength} of node $i$, defined as
\begin{equation}
S_\mathrm{OUT}(i)=\displaystyle\sum_{j=1}^{M_z} \textbf{P}_{ij} \ ,
\label{eq:Sout}
\end{equation}
can be used to formulate a general definition of the bilayer
transport matrix, which is row-stochastic and valid for both
closed and open flows (i.e. cases in which the collecting layer
is always reached and cases in which it is not):
\begin{equation}
\textbf{Q}_{ij}= \left \{ \begin{matrix} \frac{\textbf{P}_{ij}}{S_\mathrm{OUT}(i)}
& \mbox{if }S_\mathrm{OUT}(i) \ne 0\mbox{ }
\\ 0 & \mbox{if } S_\mathrm{OUT}(i)=0\mbox{ }\end{matrix}\right.  \ .
\label{eq:Qij}
\end{equation}

In the following we indicate some relevant network
measures that can be computed from this bipartite transport matrix.

\subsection{Network measures}

Many quantities have been introduced to characterize the
topology and connectivity properties of networks \cite{newman2012networks}. In this paper we
will not consider non-local quantifiers, such as optimal paths,
betweenness or communities
\cite{bolltbook2013,SerGiacomi2015,SerGiacomiMost2015,SerGiacomiDominant2015,SerGiacomi2021}.
We just introduce the simplest quantifiers involving single
nodes, namely degrees and network entropy. The adjacency matrix
is given by
\begin{equation}
A_{ij}= \left \{ \begin{matrix} 1 & \mbox{if } \textbf{Q}_{ij} > 0\mbox{ }
\\ 0 & \mbox{if } \textbf{Q}_{ij} = 0\mbox{ }\end{matrix}\right.  \ .
\label{eq:adjacency}
\end{equation}
It is used to define the out-degree of a node $i$,
$K_\mathrm{OUT}(i)$, i.e, the number of nodes in layer $z$ receiving
fluid from node $i$ in layer $z_0$; and the in-degree for a
node $j$, $K_\mathrm{IN}(j)$, which is the number of nodes of the
release layer from which fluid content arrives at node $j$ in
the collecting layer:
\begin{eqnarray}
\label{eq:degrees}
K_\mathrm{OUT}(i)&=&\displaystyle\sum_{j=1}^{M_z} A_{ij}, \\
K_\mathrm{IN}(j)&=&\displaystyle\sum_{i=1}^{M_0} A_{ij}.
\end{eqnarray}

Quantities related to degrees, but that take into account the
actual proportion of particles arriving at each node (the
\textit{weights} of the links) are the out-strength defined in
Eq. (\ref{eq:Sout}) and the \textit{in-strength}:
\begin{equation}
S_\mathrm{IN}(j)=\displaystyle\sum_{i=1}^{M_0} \textbf{P}_{ij} \ .
\label{eq:Sin}
\end{equation}
An alternative to $S_\mathrm{IN}$ can also be defined by using
$\textbf{Q}_{ij}$ instead of $\textbf{P}_{ij}$. It coincides
with (\ref{eq:Sin}) for closed flows, which is the case for the
example presented later in this paper.

Another quantity that takes into account the weights of the
links is the network entropy, defined for each node $i$ of the
release layer as
\begin{equation}
H(i)=-\displaystyle\sum_{j=1}^{M_z}\textbf{Q}_{ij}\log(\textbf{Q}_{ij}).
\label{eq:Hi}
\end{equation}
Note that, at difference with previous references
\cite{SerGiacomi2015},
we have not introduced a
prefactor corresponding to the inverse of the integration time
in the definition (\ref{eq:Hi}).

\section{Relationship between geometric and network characterization}
\label{Sec:relation}

For clarity, in the following we write expressions in terms of the
matrix $\textbf{P}_{ij}$, with the understanding that
$\textbf{Q}_{ij}$ should be used instead if the flow
is open. We first obtain a relationship between the
probabilistic or network approach and the geometric or
dynamical one for the evolution of densities. Recall that
$\textbf{P}_{ij}$ is estimated as $\textbf{P}_{ij}=N_{ij}/N_0$,
where $N_{ij}$ is the number of particles released from box
$A_i$ and landing on box $B_j$, provided $N_0$ particles are
seeded from each release box (giving the same density
$\sigma_0$ at each initial box if all of them have the same area).
Then, $S_\mathrm{IN}(j)$, defined in
(\ref{eq:Sin}), is estimated as $S_\mathrm{IN}(j) = N_j/N_0$, where
$N_j$ is the number of particles landing on box $B_j$
irrespective of their origin. On the other hand, the average of
the ratios of local densities $\sigma(\bx)/\sigma_0$ of the
points inside a collecting box $B_j$,
$\langle\sigma(\bx)\rangle_{B_j}=\mu(B_j)^{-1}\int_{B_j} d\bx
\sigma(\bx)/\sigma_0$ is also estimated by $N_j/N_0$. These estimates become exact in the limit $N_0 \to \infty$. Using
relationships (\ref{eq:bartau_barlambda}) and (\ref{eq:df}) we
find
\begin{equation}
S_\mathrm{IN}(j) = \lim_{N_0 \to \infty} \frac{N_j}{N_0} = \langle F \rangle_{B_{j}} =
\langle ( \bar{\Lambda}_{1} \bar{\Lambda}_{2})^{-1} \rangle_{B_{j}} \ ,
\label{eq:SIlambdas}
\end{equation}
where the left-hand side is computed from the network approach
of Sect. \ref{Sec:Net}, and the right-hand average is a
coarse-graining of quantities from the geometrically based
approach of Sect. \ref{Sec:Falling}. Note that Eq.
(\ref{eq:df}) assumes the absence of folding processes
producing multiple branches of arrival of the release layer
onto the collecting one, so that this is also needed for the
validity of (\ref{eq:SIlambdas}).

We now suggest some network-geometric relationships similar to
the ones developed in \cite{SerGiacomi2015} for single-layer
Lagrangian flow networks. In particular, relationships between
degree and network entropy on the one hand and the largest
stretching factor and Lyapunov exponent on the other were
found. These relationships were not exact ones, but approximate
relationships that were checked to hold for the case of long
times, sufficiently small network boxes, and a clear hyperbolic
situation (i.e. Lyapunov exponents sufficiently larger or
smaller than zero).

By repeating the heuristic arguments developed in
\cite{SerGiacomi2015} we can find the following approximate
relationships between the network and the geometrical
description of our two-layer dynamics:
\begin{eqnarray}
K_\mathrm{OUT}(i) &\approx& \left< \bar \Lambda \right>_{A_i} =
\left< e^{|z-z_0|\bar\lambda} \right>_{A_i}
\label{eq:KoutLambda}\\
H(i) &\approx& \left< \log \bar \Lambda \right>_{A_i} =
|z-z_0|\left< \bar \lambda\right>_{A_i} \ ,
\label{eq:Hlambda}
\end{eqnarray}
where $\bar\Lambda$ and $\bar\lambda$ are defined below. The averages perform a coarse-graining of the values of
$\bar\Lambda(\bx_0)$ or $\bar\lambda(\bx_0)$ over all initial
conditions inside the initial box $A_i$. At difference with
the bidimensional situation considered in
\cite{SerGiacomi2015}, in which only one of the stretching
factors was larger than one (a single expanding direction), in
the present three-dimensional dynamics several directions can be
expanding, and these directions are, in the arguments leading to
Eqs. (\ref{eq:KoutLambda}-\ref{eq:Hlambda}), the ones that
contribute to the out degree $K_\mathrm{OUT}$ or to the network
entropy $H$. In consequence, in Eqs.
(\ref{eq:KoutLambda}-\ref{eq:Hlambda}) we should use for every
initial location $\bar\Lambda \equiv \prod_\alpha
\bar\Lambda_\alpha$, where the product is over all factors $\Lambda_\alpha$
that satisfy $\Lambda_\alpha>1$ at that point. Or,
equivalently, $\bar\lambda \equiv
\sum_\alpha\bar\lambda_\alpha$, where the sum is over all
positive FDLEs, $\bar\lambda_\alpha>0$, at that point.

We stress that relationships
(\ref{eq:KoutLambda}-\ref{eq:Hlambda}) are not exact, but we
expect them to be satisfied for sufficiently small network
boxes, large $|z-z_0|$, and dynamics sufficiently hyperbolic,
which roughly requires $\bar\Lambda_\alpha$ sufficiently
different from unity. We will check this validity for a
particular flow model in Sect. \ref{subsect:numrelationship}.

\section{Numerical results}
\label{Sec:num}

In this section we illustrate the previous concepts with a
slightly modified version of an idealized incompressible 3d
flow, the ABC flow.

\subsection{ABC flow model}

The ABC flow is a 3d model flow which is widely used for
analyzing chaotic transport \cite{doi:10.1137/16M1059059,dombre_frisch_greene_henon_mehr_soward_1986}.
It provides a simple stationary solution of Euler's equation
for incompressible, inviscid fluid flows.

To simulate the situation of particles going from one layer to
another, we modify the ABC flow with a drift in a preferential
direction, specifically in the vertical one (z-direction),
without changing most of the properties of the flow. The
motivation for this choice is to mimic in a very simple way the
transport of particles falling under gravity in a chaotic fluid
flow. The equations describing the model are
\begin{eqnarray}
 \dot{x} &=& v_x = A\sin z + C\cos y, \\
 \dot{y} &=& v_y = B\sin x + A\cos z, \\
 \dot{z} &=& v_z = C\sin y + B\cos x+ D .
\end{eqnarray}
We take $A=1, B=\sqrt{2},C=\sqrt {3}$ for which chaotic motion
is found \cite{dombre_frisch_greene_henon_mehr_soward_1986}. The new constant
$D=-3.15$ is the one giving a contribution to the velocity
pointing downwards. Its value is just sufficient to keep the
particles to travel downwards in the $z$ direction (thus,
$v_z<0$ for any particle at any time). Among other
consequences, this guaranties that all initially released
trajectories will reach the collecting layer at some time, so
that $S_\mathrm{OUT}=1$ in Eq. (\ref{eq:SIlambdas}). In the horizontal
coordinates the fluid domain is $x,y \in [0,2\pi]$ with
periodic boundary conditions. In the vertical (z-coordinate)
particles are released from the layer $z_0=10$ and are followed
until they reach the layer at coordinate $z$ where integration
is stopped. Thus the model is defined in the vertical interval
$[z,z_{0}]$.


Note that $\nabla\cdot\bv=0 $.  The facts that $v_z<0$ and that
the flow is time-independent guarantee that the map
$\phi_{z_0}^z$ is one-to-one.

\subsection{Transport properties between layers}
\label{subsect:numtransport}

\begin{figure}[h]
	\includegraphics[scale=0.3]{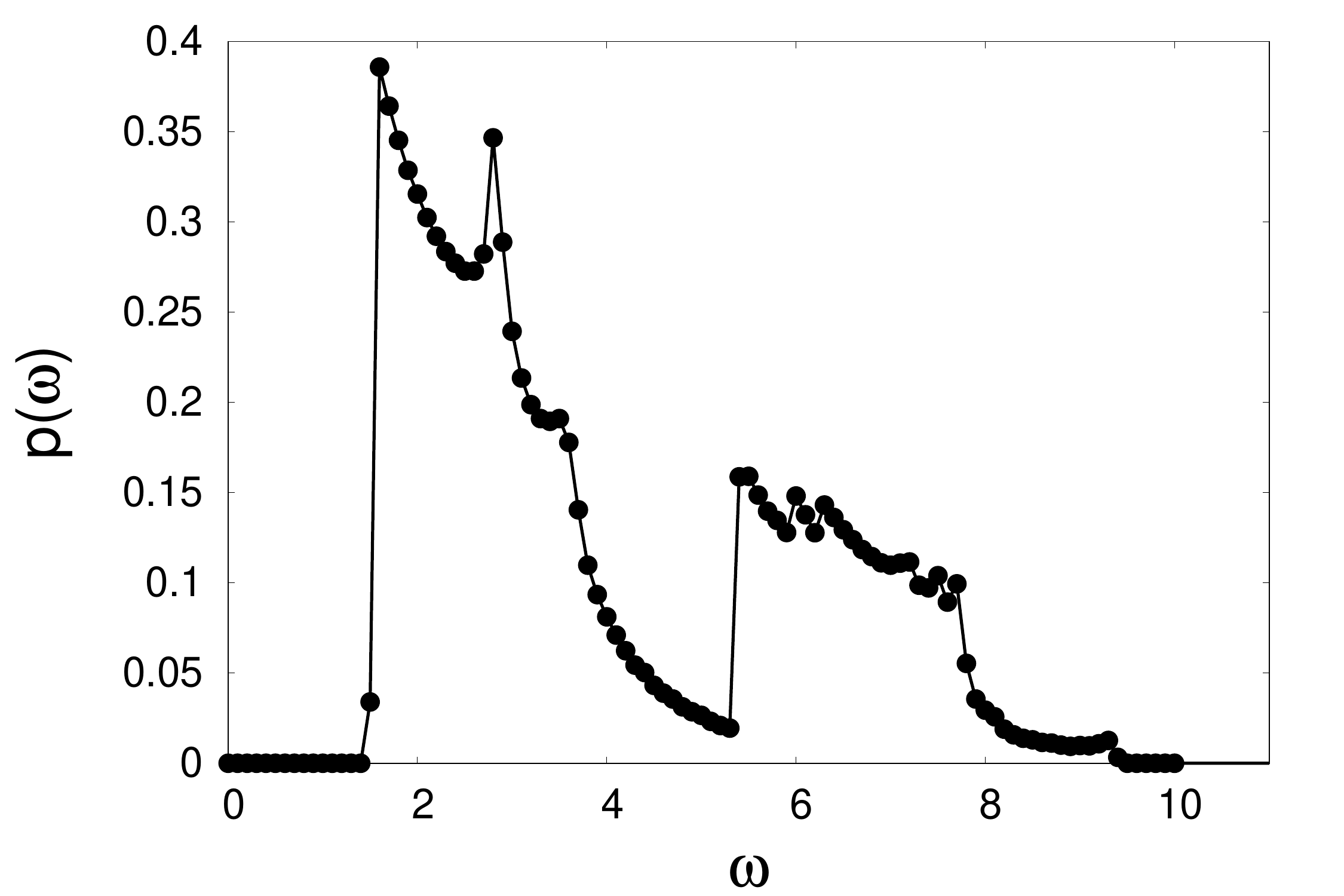}
	\caption{Histogram of the times to reach the second layer at $z=0$, starting from
	$z_0=10$.}
	\label{fig:times} 
\end{figure}

\begin{figure}[h]
\begin{center}
\includegraphics[scale=0.8]{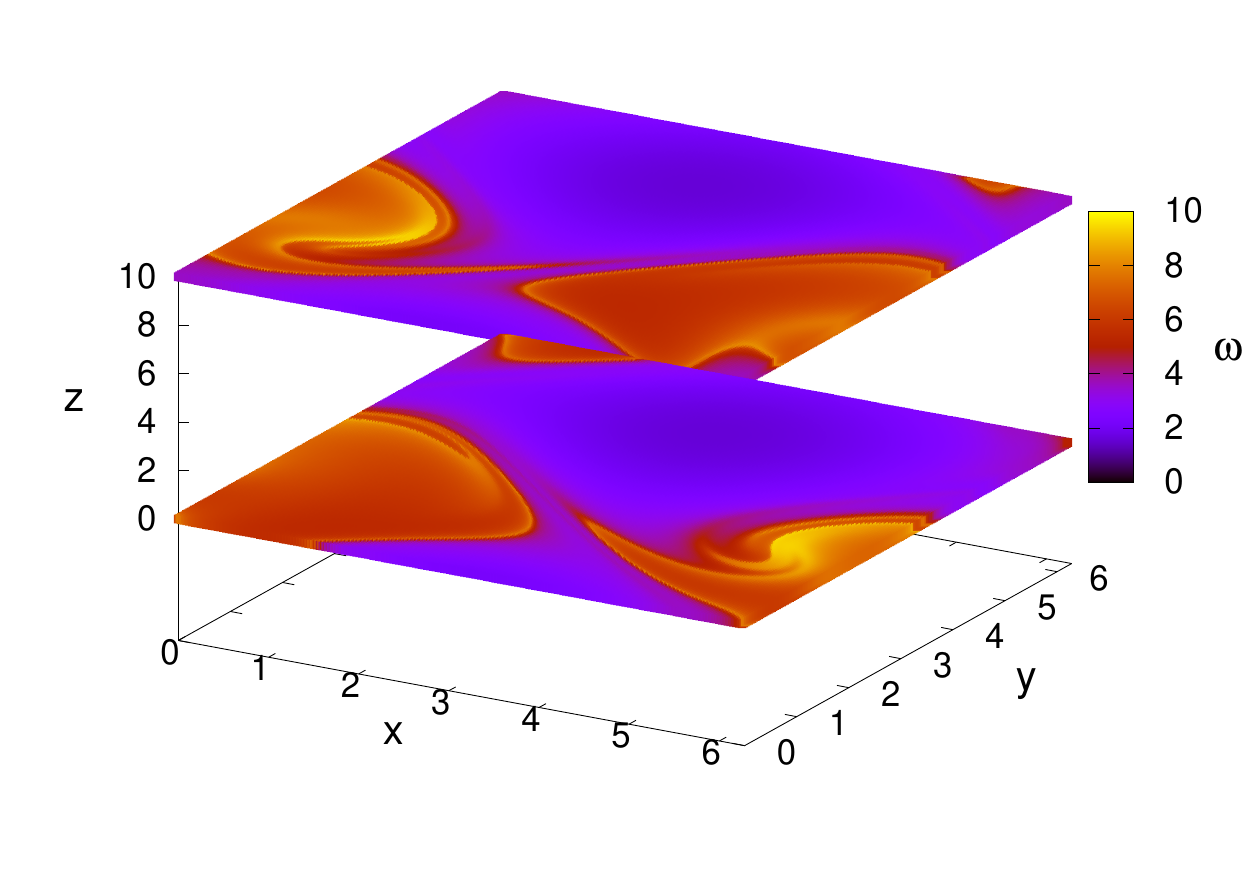}
\caption{The travel time, $\omega$, from release layer
($z_0=10$) to collecting layer ($z=0$) displayed at
the initial and final position of each particle.}
\label{fig:tiempomapping}
\end{center}
\end{figure}

We first study the map $\phi_{z_{0}}^{z}$ for the ABC flow by
taking $z_{0}=10$ and $z=0$ (particles fall from height $z_0$).
In Fig. \ref{fig:times} we show a histogram of arrival times,
$p(\omega)$. It  shows a two-peaked shape with peaks around the
values $2$ and $6$. We can differentiate two main dynamical
behaviors: more laminar for the first peak and more chaotic for
the second one. This suggests the existence of two zones of
trajectory behavior in the fluid flow, which is confirmed in
Figure \ref{fig:tiempomapping}.

We show in Figure \ref{fig:tiempomapping} the spatial
distribution of $\omega$, the time needed by every particle to
go from layer $z_0$ to layer $z$. This time is shown as a color
map for every particle at the release layer $z_0$ and on its
corresponding final position at layer $z$. The color map in the
bottom layer is conveniently computed by running the flow
backwards in time from a regular grid of initial conditions
located at $z$. The equivalence between the backwards- and the
forward-in-time calculation of $\omega$ is guaranteed by the fact
that for this time-independent flow the map $\phi_{z_0}^z$ is
one-to-one. Since $v_z<0$ for any particle and time, all
particles released in the upper layer reach the collecting
layer in a finite time, and all locations in the collecting
layer receive a trajectory.

We observe the two regions in Fig. \ref{fig:tiempomapping}: the first peak in Figure
\ref{fig:times} corresponds the dark regions, with more laminar
trajectories, i.e., the particles go straightforwardly from one
layer to the other; and the red regions correspond to the
second peak and to more convoluted (chaotic) trajectories. The
frontiers between initial conditions of large and small $\omega$
are quite sharp, and will be identified with lines of large
finite-depth Lyapunov exponent in Section
\ref{subsect:numlyapunov}.

\subsection{Geometric characterization}
 \label{subsect:numlyapunov}

The Jacobian ${\bm{\bar{J}}}_\mathcal{M}$ is computed by releasing
particles on a regular grid on layer $z_0$, integrating their
trajectories under the modified ABC flow until reaching the
final layer at $z$, and approximating the derivatives in ${\bm{\bar{J}}}_\mathcal{M}=\nabla\phi_{z_0}^z (\bx_0)$ by finite differences
between final positions of initially neighboring particles.
Then, its singular values $\bar \Lambda_1$ and $\bar \Lambda_2$
are computed after construction of the Cauchy-Green tensor
$\bar C_\mathcal{M}={\bm{\bar{J}}}_\mathcal{M}^T {\bm{\bar{J}}}_\mathcal{M}$.
\begin{figure}[h]
\includegraphics[width=.8\columnwidth]{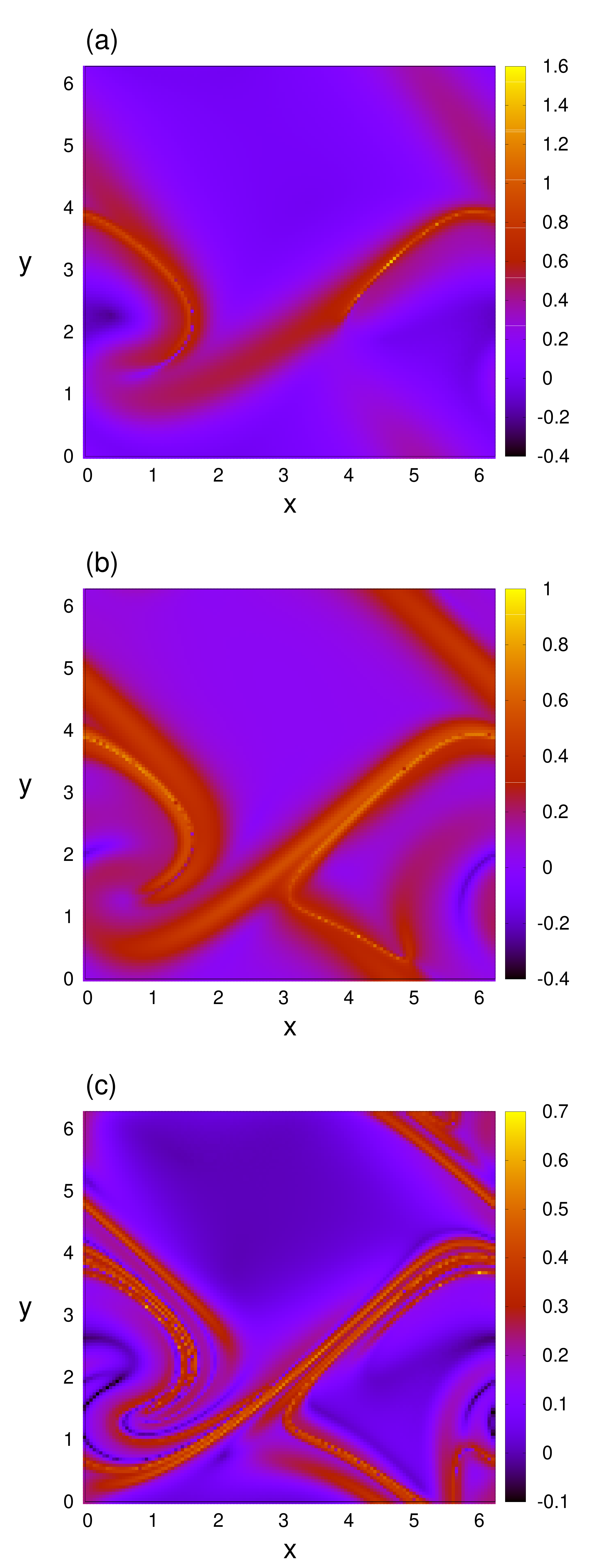}
	\caption{Maximal FDLE $\bar\lambda_1$ for dynamics under the modified ABC flow, displayed at
the initial particle locations in the release layer $z_{0}=10$, and for collecting layer
at (a) $z=2\pi$, (b) $z=4$ and (c) $z=0$.}
	\label{fig:maxlyap}
\end{figure}

Figure \ref{fig:maxlyap} shows the maximal FDLE
$\lambda_1(\bx_0)$ from Eq. (\ref{eq:FDLE}), displayed on the
release layer $z_0=10$, for collecting layers at three
different depths $z$. We see that increasingly finer
filamentary structures appear for increasing travel depth. This
is similar to the behavior of the FTLE for increasing
integration time. We note that the highest FDLE values roughly divide
the release domain into two regions (remember the periodic
boundary conditions in the horizontal directions) that closely
correspond to the long and short travel time regions in Fig.
\ref{fig:tiempomapping}: as for the FTLE, ridges of FDLE are
associated with separatrices that divide the release layer into
regions of different dynamic behavior.
In particular, these structures are reminiscent of a stable
 foliation corresponding to hyperbolic
trajectories. Although periodic trajectories cannot exist when
$v_z < 0$ everywhere in a domain with a finite vertical
extension at any time, they can exist in the same velocity
field with periodic boundary conditions in the vertical
direction. Finite portions of such trajectories will govern
finite-time chaotic dynamics through finite-length versions of
the corresponding stable and unstable manifolds that appear
according to the extent of the domain in the $z$ direction when
periodicity is not prescribed for that coordinate. Ridges in
the FTLE field would arise from intersections with the release
layer of these finite-length stable manifolds, and this also
happens at the same locations in the FDLE field, as seen in
Fig. \ref{fig:maxlyap}, in spite of the complication that
arises from the projection effect included in the definition of
the FDLE. In fact, we have checked (not shown) that these
intersections are much more clearly identified in the FDLEs than in
the FTLEs. 

\begin{figure}[h]
	\includegraphics[width=0.8\columnwidth]{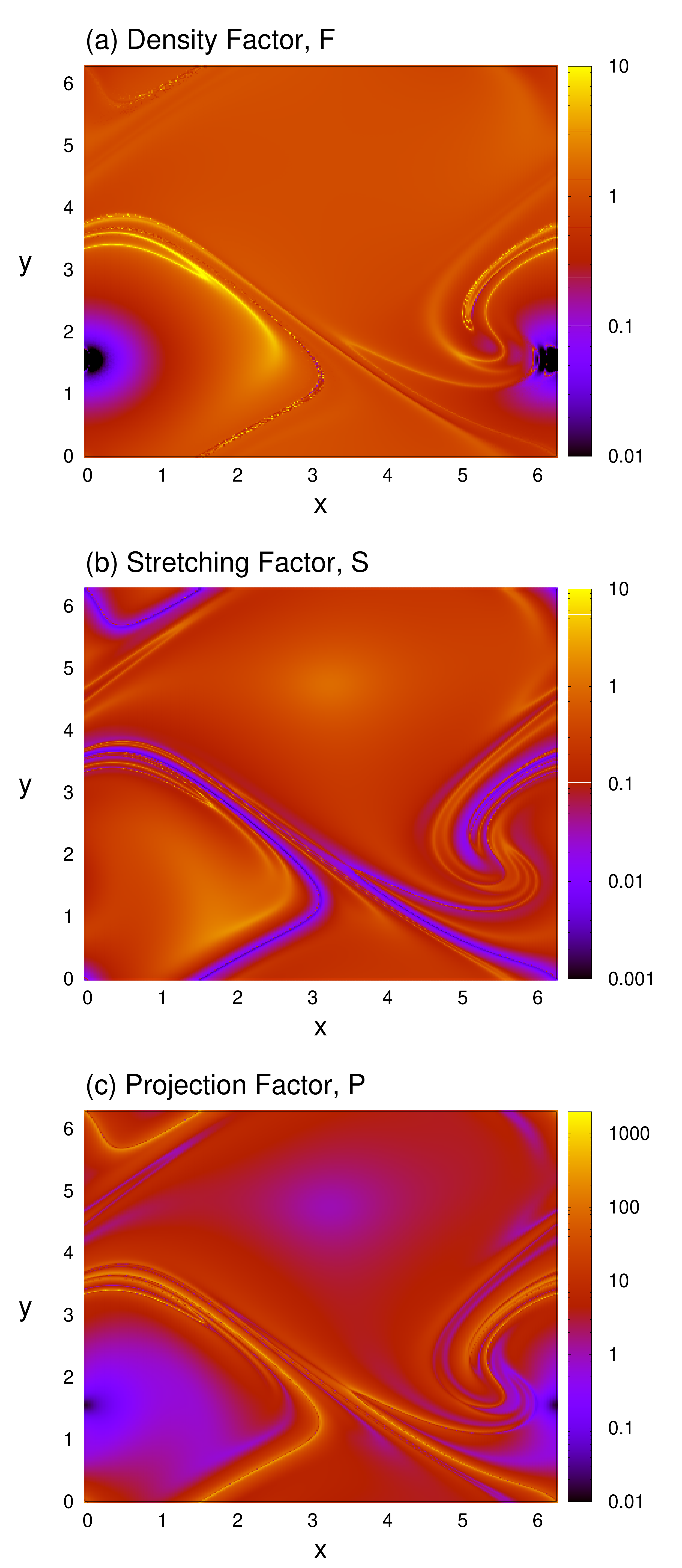}	
	\caption{(a) The density factor, $F$, computed
as $F=(\bar\Lambda_1\bar\Lambda_2)^{-1}$
at the collecting layer $z=0$, giving the relative density of
collected particles if the density in the release layer $z_0=10$ is uniform. Panel (b) shows the
stretching factor $S$, computed as $S=(\Lambda_1\Lambda_2)^{-1}$. Panel
(c) shows the projection factor $P$ from Eq. (\ref{eq:SyP2}). We have checked that
$F=S~P$ to good accuracy. Note the logarithmic scale in the color maps.}
	\label{fig:F}
\end{figure}

It also appears that there is a correspondence
between the intersections with the collecting layer of
finite-length unstable manifolds and ridges in the density
factor $F$: in Figure \ref{fig:F}a we plot the factor $F$ on
the collecting layer, which is the factor that multiplies the
initial density at the release layer (and thus it is
proportional to the accumulated density of particles if the
release density is constant). We also display in the other
panels of Fig. \ref{fig:F} the two geometric factors,
stretching $S$ and projection $P$, that shape $F$ (i.e.
$F=S~P$, Eq. (\ref{eq:eqF})). We see clearly from the plot of
$F$ that filamentary structures will appear in the density
collected in the lower layer. The effect of surface-element
stretching ($S$) is less determinant for $F$ than the
projection of surface elements onto the collecting layer, $P$,
although this can be different for other types of flows. In
more complex flows \cite{Drotos2019,Monroy2019, Sozza2020} the
projection factor can even diverge at {\sl caustics}, locations
where the denominator of Eq. (\ref{eq:SyP2}) vanishes. As in
\cite{Drotos2019}, there is some degree of anticorrelation
between $S$ and $P$, so that the fluctuations in $F$ are
smaller than those in $S$ and $P$.

\subsection{Network characterization}
\label{subsect:numnet}

We study connectivity properties between layers $z_0=10$ and
$z=0$. For doing this, we divide the upper layer into $100
\times 100$ square boxes $A_{i}, \ i=1,...,10000$, and the
lower one into $100 \times 100$ square boxes $B_{j}, \
j=1,...,10000$. Then we release from each box in $z_0$ $900$
particles uniformly distributed. We integrate each of these
particles with the map $\phi_{z_{0}}^{z}$ (equivalent to
integrating Eq. (\ref{eq:lag}) until reaching the collecting
layer at $z$).

\begin{figure}[h]
	\includegraphics[scale=0.28]{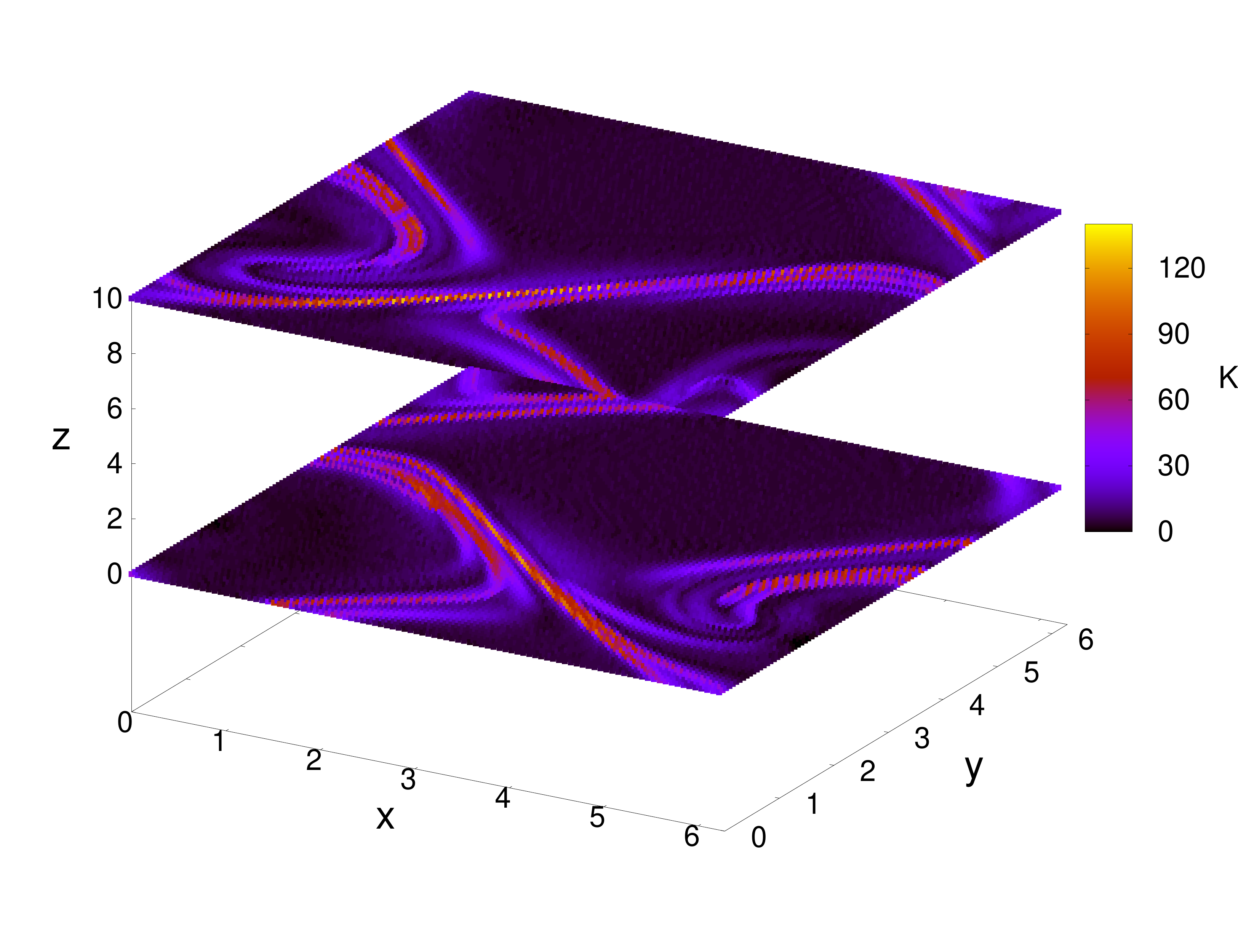}
	\caption{Out-degree and in-degree in the release ($z_0=10$) and the arrival ($z=0$)
layers, respectively.}
\label{fig:degrees}
\end{figure}

In Figure \ref{fig:degrees} we show the out-degree in the
starting layer and the in-degree in the final one. The
out-degree for a given box in the starting layer indicates the
number of boxes reached in the final layer. It is a measure of
dispersion, and large values at a box indicate that a part of a
repelling or dispersing structure is present there. On the
other side, large values of in-degree in the final layer
indicate mixing from a large number of different initial
conditions, so that boxes with in-degree maxima trace the
location of attracting regions. These ideas are confirmed when
comparing degrees to the FDLEs
of Fig. \ref{fig:maxlyap}c.
\begin{figure}[h]
	\includegraphics[width=\columnwidth]{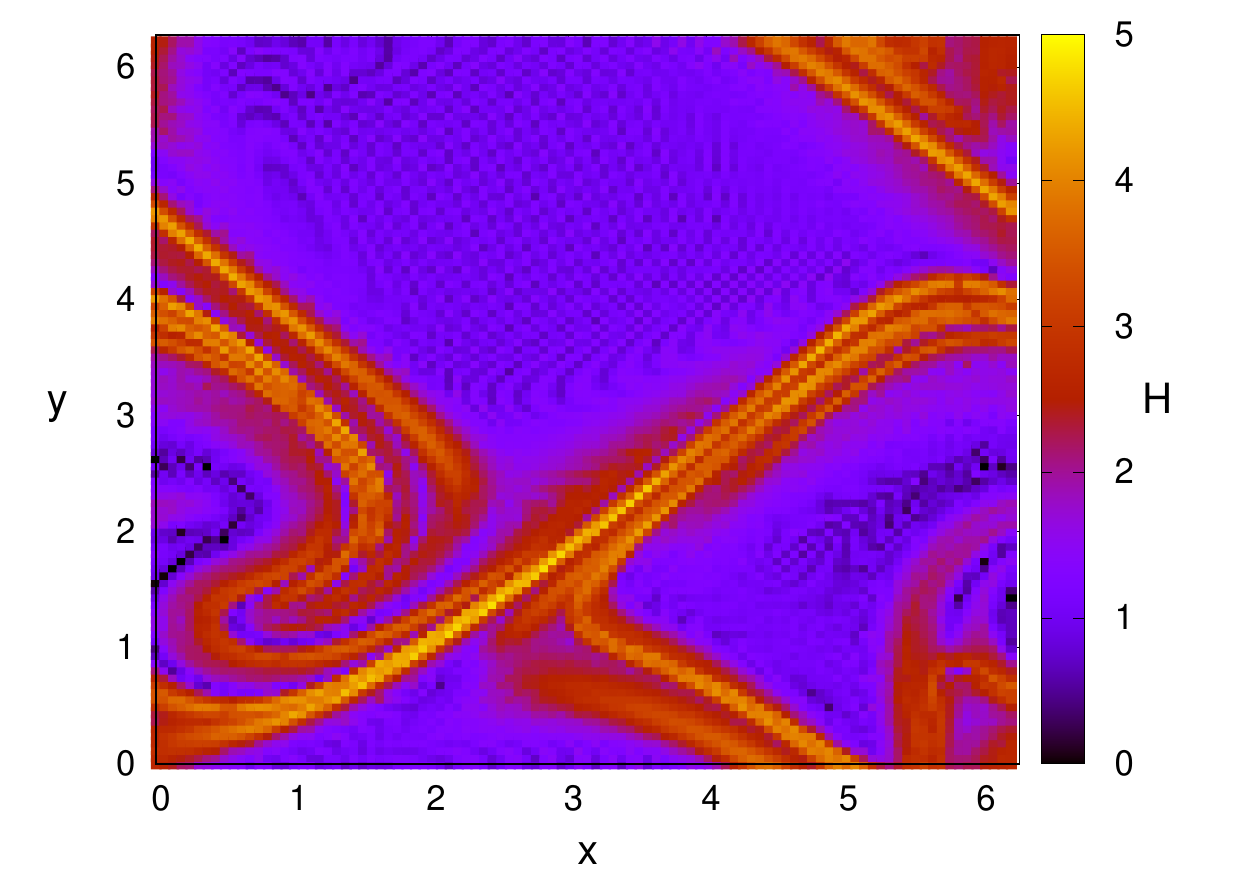}
	\caption{Entropy $H(i)$ for transport from the release layer at $z_0=0$ to the collecting layer
at $z=0$, displayed on the release layer.}
\label{fig:H}
\end{figure}

Another quantity computed in the network approach, the entropy
$H(i)$ defined in Eq. (\ref{eq:Hi}) is displayed in Fig.
\ref{fig:H}. There is a clear relationship with $K_\mathrm{OUT}(i)$
(Fig. \ref{fig:degrees}), and also with the FDLEs
of Fig. \ref{fig:maxlyap}c. These
relationships will be checked more systematically in the next
section.

\subsection{Relationship between geometric and network
characterization}
\label{subsect:numrelationship}

\begin{figure}[h]
	\includegraphics[width=0.8\columnwidth]{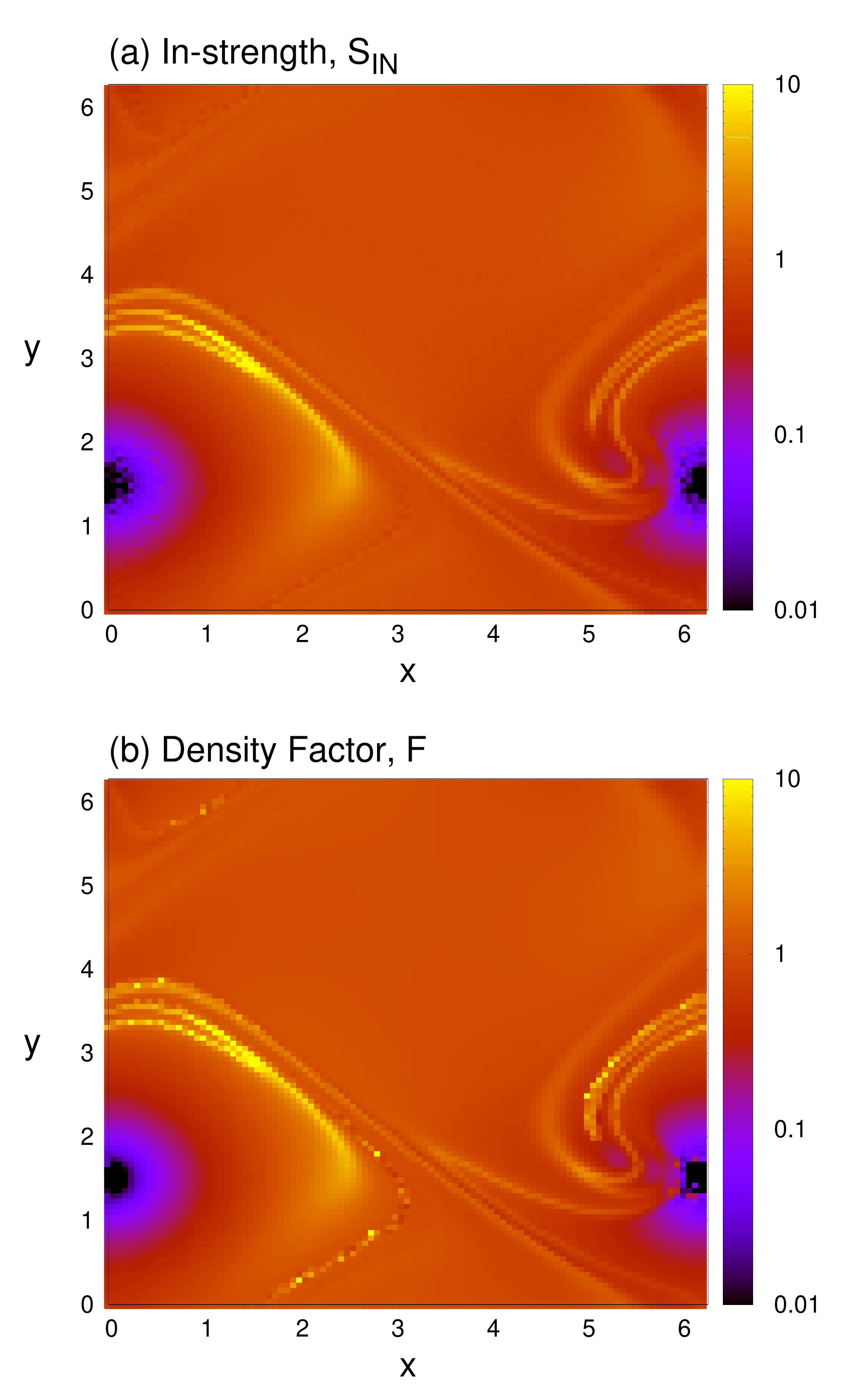}
	\caption{(a) The in-strength $S_\mathrm{IN}(j)$ in the lower layer $z=0$, which gives the accumulated density in that layer starting from a unit-density uniform release at $z_0=10$. (b) Density factor averaged on each box
of the accumulation layer, i.e. $\langle F\rangle_{B_{j}}= \langle (\bar{\Lambda}_{1} \bar{\Lambda}_{2})^{-1} \rangle_{B_{j}}$. }
\label{fig:densidad}
\end{figure}

In this section we first check Eq. (\ref{eq:SIlambdas}). It
relates the network quantity $S_\mathrm{IN}(j)$, giving also the
density accumulated at box $B_j$ in the lower layer relative to
the uniform release density in the upper layer, to a coarse
graining on collecting boxes of a quantity developed in the
geometric approach, the density factor
$F=(\bar\Lambda_1\bar\Lambda_2)^{-1}$. In Fig.
\ref{fig:densidad} we see that, as predicted, both quantities
are nearly equal, although there are some differences in the
narrowest filamental regions, arising from numerical
inaccuracies. According to the outlying values in Fig. \ref{fig:densidad}b, it is presumably $S_\mathrm{IN}$ that can be computed more reliably than $(\bar\Lambda_1\bar\Lambda_2)^{-1}$.





We now address the validity of expressions
(\ref{eq:KoutLambda}) and (\ref{eq:Hlambda}). At difference
with Eq. (\ref{eq:SIlambdas}), these formulae were
derived only heuristically, following the arguments
of Ref. \cite{SerGiacomi2015}. Their validity is subjected to
restrictions such as smallness of boxes, large values of
$|z-z_0|$, and sufficiently hyperbolic dynamics (roughly,
singular values sufficiently different from unity), which we
will now check if are satisfied for our modified ABC flow.
\begin{figure}[h]
	\includegraphics[width=0.8\columnwidth]{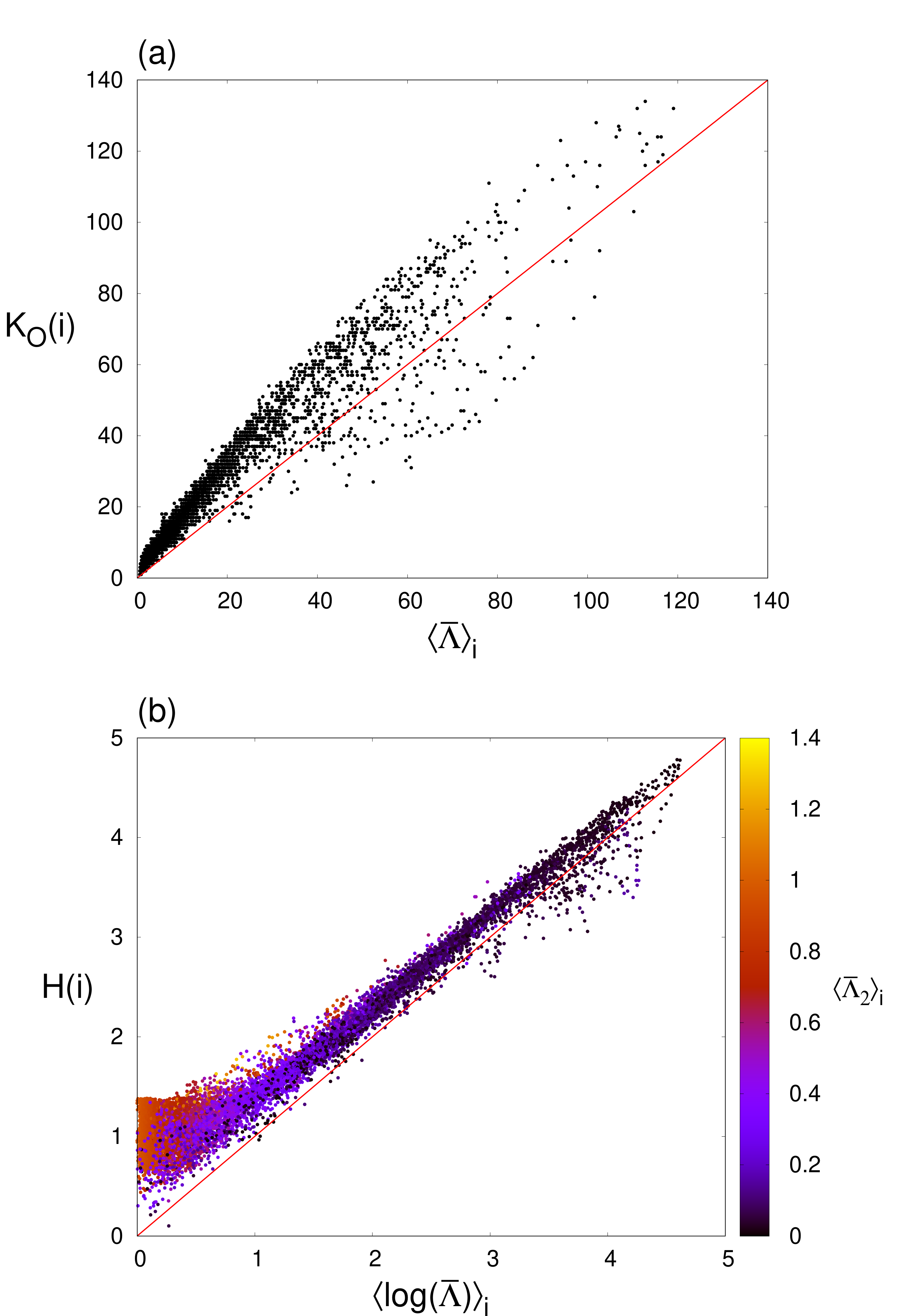}
	\caption{(a) Scatter plot of values of $K_\mathrm{OUT}(i)$ vs $\langle \bar\Lambda \rangle_{A_i}$. The red diagonal indicates the fulfillment of Eq. (\ref{eq:KoutLambda}). (b) Scatter plot of values
of $H(i)$ vs $\langle \log\bar\Lambda \rangle_{A_i}$. The red diagonal indicates
the fulfillment of Eq. (\ref{eq:Hlambda}). Dots are colored according to the value of
$\langle \bar\Lambda_2 \rangle_{A_i}$.}
\label{fig:hksvslyap}
\end{figure}

\begin{figure}[h]
	\includegraphics[width=\columnwidth]{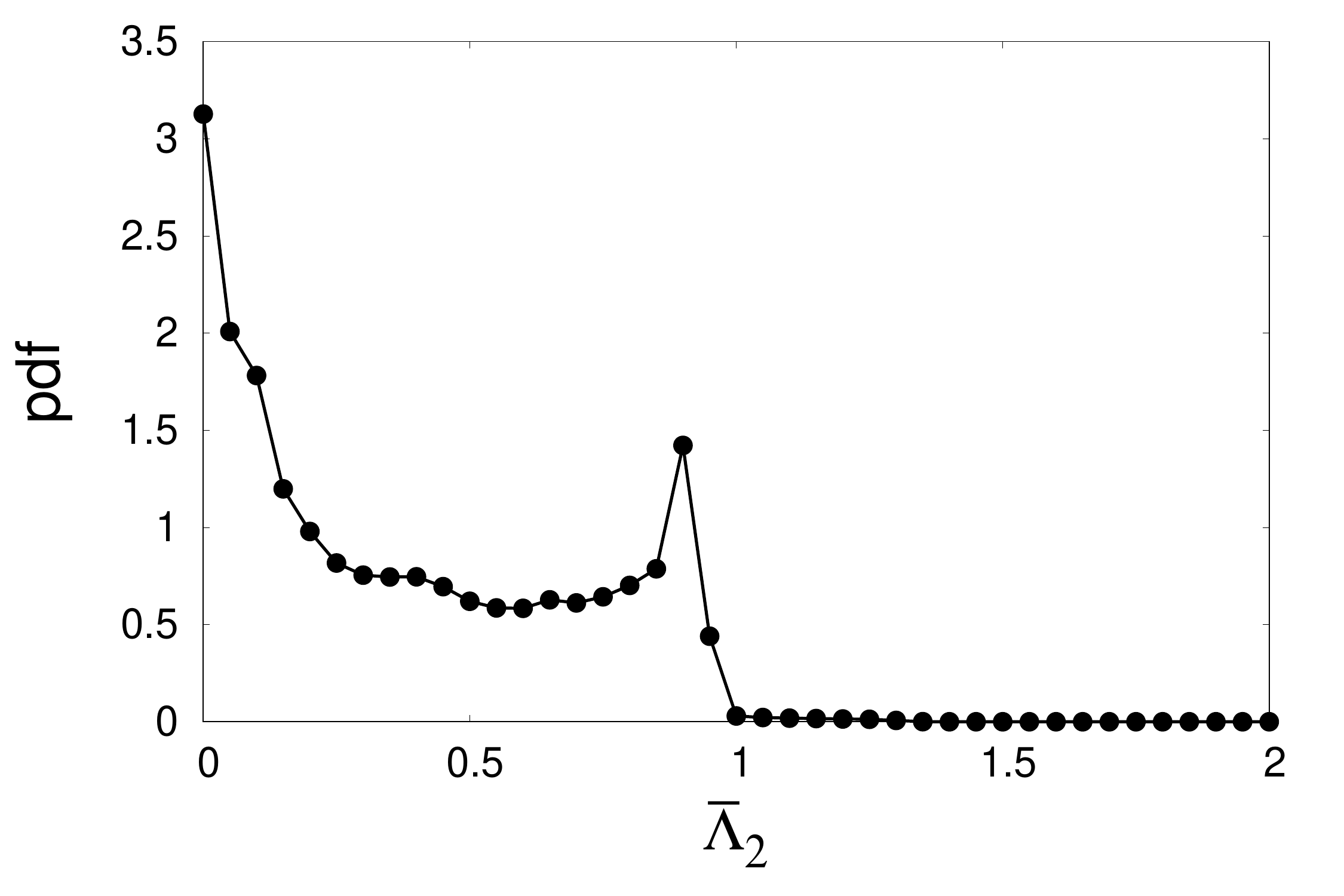}
	\caption{Probability density function of the values of
$\bar{\Lambda}_{2}$ on the release layer.}
\label{fig:eig2}
\end{figure}

Regarding Eq. (\ref{eq:KoutLambda}), comparison of $K_\mathrm{OUT}$
from the upper layer in Fig. \ref{fig:degrees}, and
$\bar\lambda_1$ in Fig. \ref{fig:maxlyap}c, which
is the logarithm of $\bar\Lambda_1$, already indicates a strong
relationship. A more quantitative comparison is made in panel
(a) of Fig. \ref{fig:hksvslyap} between $K_\mathrm{OUT}(i)$ and
$\langle \bar\Lambda \rangle_{A_i}$, where
$\Lambda=\prod_\alpha\bar\Lambda_\alpha$
is the product of all singular values larger than unity.
We see that,
although there is a positive correlation, there is no identity
between the two quantities. We attribute this failure of Eq.
(\ref{eq:KoutLambda}) to the fact that the second singular
value $\bar\Lambda_2$ takes values close to unity for most of
the trajectories. This is confirmed by the distribution of
$\bar\Lambda_2$ in the upper layer displayed in Fig.
\ref{fig:eig2}. We note that, since the modified ABC flow is
time independent, we have always that the second Lyapunov
exponent is zero, or $S_2=1$. The three-dimensional singular
value $S_2$ is not exactly $\Lambda_2$ nor $\bar\Lambda_2$, but
it is related to them at long times, which justifies the
prevalence of values $\bar\Lambda_2\approx 1$ in Fig.
\ref{fig:eig2}, and then a lack of hyperbolicity.
$\bar\Lambda_2 \approx 1$ implies that boxes in the upper layer
are not converted by the dynamics into thin filaments, but into
broad strips. When reaching the collecting layer, they will
leave a footprint larger than the thin filament needed to
derive Eq. (\ref{eq:KoutLambda}), and consequently $K_\mathrm{OUT}$
will be generally larger than predicted, as seen in Fig.
\ref{fig:hksvslyap}a.

Relationships that imply a weighting with the number of
particles reaching a particular box in the collecting layer are
expected to be more robust than relations such as Eq.
(\ref{eq:KoutLambda}) that involve the degree, a quantity
counting all boxes to which particles arrive, independently on
how many of them do so. Thus, Eq. (\ref{eq:Hlambda}), although
derived under heuristic arguments similar to those leading to
Eq. (\ref{eq:KoutLambda}), is expected to be satisfied under a
broader range of conditions. This is indeed the case, as seen
by comparing plots of entropy (Fig. \ref{fig:H}) with
corresponding plots of FDLE (Fig.
\ref{fig:maxlyap}c). A more quantitative check is performed in
panel (b) of Fig. \ref{fig:hksvslyap}. We see that the equality
in Eq. (\ref{eq:Hlambda}) is satisfied much better than Eq.
(\ref{eq:KoutLambda}). Nevertheless, there are still deviations,
especially for small values of $\langle \log\bar\Lambda
\rangle_{A_i}$. These small values arise from locations where
$\bar\Lambda\approx 1$, confirming situations of lack of
hyperbolicity. We have also colored the points in the scatter
plot with the values of $\langle\bar\Lambda_2\rangle_{A_i}$.
Again, the stronger deviations occur when both
$\langle\bar\Lambda_2\rangle_{A_i}$ and
$\langle\bar\Lambda\rangle$ are close to unity.

%


\section{Conclusions}
\label{Sec:con}

In this paper we have developed a formalism to characterize
transport of particles between two layers in a fluid. The
motivation was to obtain a theoretical framework to analyze
problems related to the sinking of particles in fluid flows,
sedimenting towards a bottom layer. Two complementary sets of
tools have been addressed: geometrical or dynamical, by
studying the dynamics and deformation of a layer of particles,
and probabilistic, using concepts from network theory. Most
importantly, we have addressed the relationship between these
two approaches, and illustrated the whole formalism with a
modified ABC model.

The crucial step is the definition of a two-layer map, which
drives particles from one initial layer to the final one.
Within the geometric approach we have analyzed the deformation
of surfaces and lines of particles released from the upper
layer. A quantity related to the
Lyapunov exponent, the FDLE, has been defined and related to
the quantities above. Within the probabilistic methodology the
natural description of the system is via bipartite networks, in
which quantities such as the out-degree in the initial layer
and the in-degree in the final one acquire a clear physical
meaning. Both descriptions have been connected, for example, by
expressing the accumulated density of particles in terms of the
in-degree and of averages of singular values defined in the
geometric approach. Other geometric-network relationships that
were successfully tested for transport on a single layer
\cite{SerGiacomi2015} are satisfied here with poor accuracy.
This stresses the need for sufficiently hyperbolic dynamics to
justify some of the heuristic steps used in the derivations.

More explicitly, the two-layer map provides a
general description of particle transport between layers,
without any restriction to hyperbolic flows or transport
without folding. This means that most of the geometric and
network formalism described in Sections \ref{Sec:Falling} and
\ref{Sec:Net}, respectively, can be applied to any type of
flow. However, some of the specific relationships we have
obtained, namely Eqs. (\ref{eq:df}) and (\ref{eq:SIlambdas}-\ref{eq:Hlambda}), require the validity of additional
hypotheses that we now detail.

The heuristic arguments leading to Eqs.
(\ref{eq:KoutLambda}-\ref{eq:Hlambda}), which link the
geometrical perspective with the network-based description, are
restricted to sufficiently hyperbolic dynamics, meaning in this
context that $\bar{\Lambda}_1$ and $\bar{\Lambda}_2$ should be
sufficiently different from unity. Thus, these two
relationships will be valid only in regions dominated by
strain. Unlike in two-dimensional incompressible flows where it
is sufficient to take care of one singular value of the
Jacobian matrix \cite{Drotos2019}, the second singular value of
$\bar{\bm{J}}_{M}$ in three-dimensional flows is independent of
the first one and thus also plays a role. If this second
singular value $\bar{\Lambda}_2$ is close to unity, fluid
patches released from the upper layer may be converted into
broad strips after being projected onto the collecting layer,
which results in a deviation from Eqs.
(\ref{eq:KoutLambda}-\ref{eq:Hlambda}).
This dependence on the second singular value
is illustrated in Fig. \ref{fig:hksvslyap}(b).

On the other hand, folding of the falling
surface, which may occur in time-dependent flows, affects our
formalism in two ways. The first is that the inverse of the
two-layer map, appearing in (\ref{eq:Pij}), is multivalued if
foldings are present, for which Eqs. (\ref{eq:df})  and
(\ref{eq:SIlambdas}) have to be modified (as done in
\cite{Drotos2019,Sozza2020}) to take into account all preimages
of each given point in the collecting layer. The second is that
the singular values of the Jacobian matrix $\bar{\bm{J}}_{M}$
are ill-defined at folds, so that the evaluation of FDLEs and
the density factor $F$ becomes impossible there as well. The
decomposition $F = S$ $P$ and the divergence of $P$ shows, in
fact, that $F$ also diverges at folds, identifying the
appearance of caustics (cf. \cite{Drotos2019}).

Note that we have assumed homogeneity in the initial
distribution of particles to focus on the effects of transport.
If one is interested in analyzing the evolution of
nonhomogeneous initial particle distributions, the density at
the collecting layer can be simply recovered by multiplying the
initial density by the corresponding density factor of each
particle trajectory reaching the bottom layer. Thus, final
densities can always be computed if the initial density of
particles at the release layer is known.

There are recent works studying, on the one side, microplankton
sedimentation in the ocean with network tools
\cite{Nooteboom2019} and, on the other, the geometry of
sedimentation dynamics and distribution of biogenic particles
\cite{Monroy2017,Monroy2019} and microplastics
\cite{delaFuente2021}. We have presented here
steps that connect both approaches, and that may provide new
insights into problems of sinking particles in the ocean. In
particular, the FDLE is a novel measure specifically defined
for the study of flow patterns between two layers with a
preferential direction of motion and quantifies structures in a
different way if compared to standard geometrical measures,
such as the classical Finite-Time Lyapunov Exponent. For
example, FDLE ridges neatly separate regions in upper and lower
layers in which particle travel times are significantly
different (compare Figs. \ref{fig:tiempomapping} and
\ref{fig:maxlyap}). Also, while the decomposition of $F$ into
$S$ and $P$ is not new in itself, we have provided here new
ways to compute $F$ and $S$. Such a decomposition is crucial
for exploring and quantifying the relative contributions of the
stretching factor $S$ and the projection factor $P$ to the
resulting distribution of particles when being collected after
a sedimentation process.
In general terms, our formalism characterizes repelling and
attracting structures associated with transport between both
layers. The result is a theoretical characterization that may
be useful in future applications that focus on transport
properties of sinking particles, such as the study of
sedimentation patterns, and barriers between regions with
qualitatively different dynamics. Furthermore, as a novel
application, community detection approaches that become
accessible thanks to the network characterization, can be
practically useful as has been the case in situations of
horizontal transport \cite{SerGiacomi2015}. 

Comparing the approach of our Section
\ref{Sec:Net} to that in \cite{Nooteboom2019} where bilayer
networks are also used, the crucial differences are that
\cite{Nooteboom2019}, using a backwards-in-time approach,
focuses on the origin over the surface of the particles
deposited on the sea-floor, and that they are interested in a
statistical description over paleo-scales. In contrast, our
network approach is based on a forward-in-time integration, so
that we focus on the fate of the particles after being released
from the surface. We thus identify flow structures at the time
scales during which the particles move from one layer to
another, and we relate them to the geometry of a falling
layer. All of this is suited to the application to mesoscale
and submesoscale transport in the marine environment, at time
scales from days to months. This will be also relevant for
studies of sedimentation 
in atmospheric flows, as for example in the context of
deposition of volcanic ashes or aerosol particles
\cite{Haszpra2011,Haszpra2019}. More generally, we expect our
formalism to be of use in other flow problems in which a
dominant direction of transport occurs.



\begin{acknowledgements}
We acknowledge MCIN/ AEI/10.13039/501100011033/ and FEDER ``Una
manera de hacer Europa" for its support to the project
MDM-2017-071, Maria de Maeztu Program for Units of Excellence
in R\&D. R.F. also acknowledges the fellowship no.
BES-2016-078416 under the FPI program of MINECO, Spain.
\end{acknowledgements}

\appendix*
\section{}
\renewcommand{\thefigure}{A\arabic{figure}}
\setcounter{figure}{0}

In this Appendix we give further details on the geometry of
projection and stretching that is used in the geometric
approach. Some of the expressions presented here were already
derived or used in Refs. \cite{Drotos2019,Monroy2019,Sozza2020}.
\begin{figure}[h]
	\includegraphics[width=.8\columnwidth]{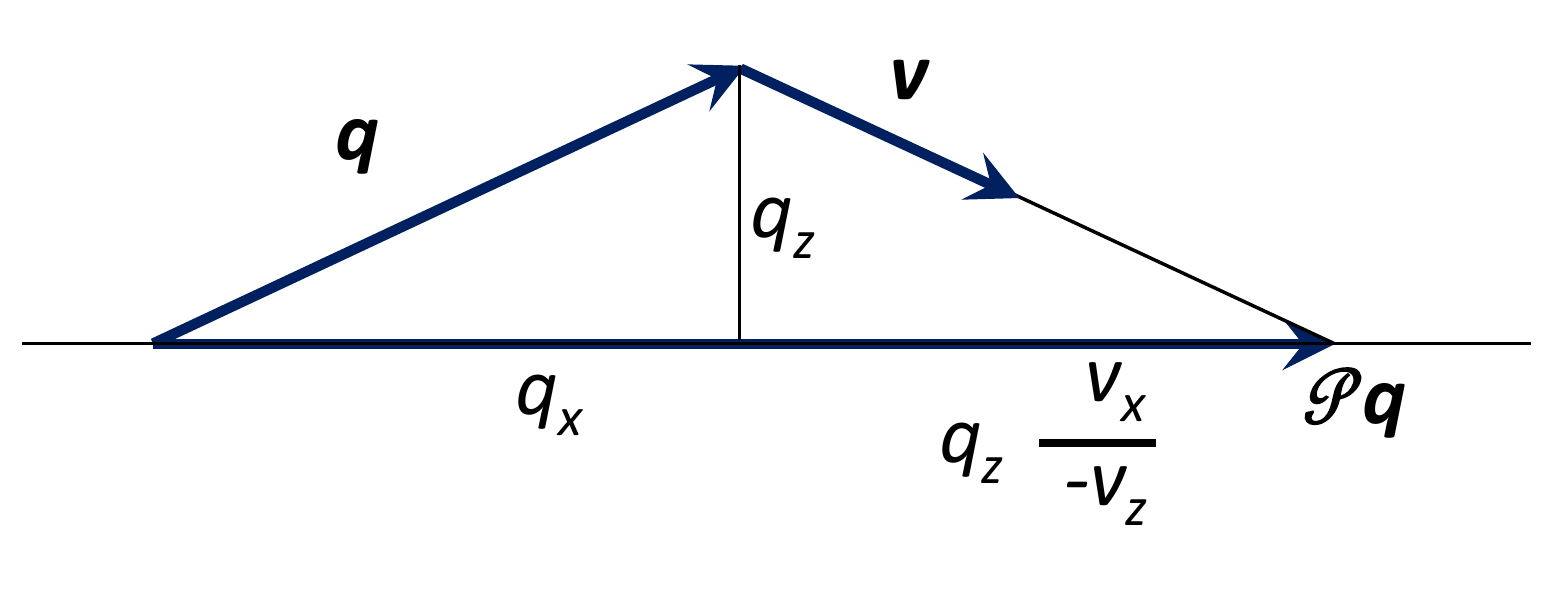}
	\caption{Sketch (in a two-dimensional situation) of the footprint or projection in the direction of
its motion, $\mathcal{P}\bq$, of a vector
$\bq=(q_x,q_y)$ onto a horizontal collecting layer (in fact a collecting line) when arriving there with
velocity $\bv=(v_x,v_y)$. We have $\mathcal{P}\bq = (q_x-q_z v_x/v_z)\hat\bx$, where $\hat\bx$ is the
unit vector in the direction of the collecting line. }
\label{fig:ProjectionDiagram}
\end{figure}

First, we derive expressions for the footprint left by a vector
$\bq$ on the collecting layer as it arrives in it with a
velocity $\bv$.
We will apply the expressions to vectors
tangent to the falling surface that represent infinitesimal
segments of that surface. 
Thus the velocity vector $\bv$ is evaluated at
the center point of the falling segment when it touches the
collecting layer at $t_z$ and, because of the segment's
infinitesimal character, the same $\bv$ applies to the whole
vector $\bq$. Figure \ref{fig:ProjectionDiagram} shows a
sketch of the geometry in a two-dimensional situation, so that
the components of the vector are $(q_x,q_z)$ and those of the
velocity $(v_x,v_z)$. The horizontal projection
$\mathcal{P}\bq$ of $\bq$ along the direction of motion, or
footprint, is made of two parts: $q_x$ and the
result of multiplying $q_z$ by the tangent of the angle between
$\bv$ and the vertical, i.e. $q_z v_x/(-v_z)$. Considering also
the $y$ component, the projected vector is
$\mathcal{P}\bq=(q_x-q_z v_x/v_z,q_y-q_z v_y/v_z)$ and the
vertical component $(\mathcal{P}q)_z$ is zero. The projection
of $\bq=(q_x,q_y,q_z)$ onto the direction of motion is a linear
operation and thus it can be expressed as the action of a
matrix $\mathcal{P}$ on the vector, with
\begin{equation}
\mathcal{P} =
\begin{pmatrix}
1 & 0 & -v_x/v_z\\
0 & 1 & -v_y/v_z
\end{pmatrix}
\ .
\end{equation}
A row of zeros can be added to the bottom if $\mathcal{P}\bq$
is considered to be embedded in three-dimensional space. An
equivalent expression for this projection operator can be
written in terms of cross products:
\begin{equation}
\mathcal{P}\bq = \hat\bz \times \left( \bq \times \frac{\bv}{v_z} \right) \ ,
\end{equation}
where $\hat\bz$ is the unit vector in the positive vertical
direction.

Let us consider a vector of the form $\bm{\tau}(t)=\frac{d
\phi_{t_0}^\tau(\bx_0(s))}{ds}$, which is tangent to the falling
surface at every time, and points initially (at time $t_0$)
along the direction on the release layer specified by the
parameter $s$.  $\bm{\bar\tau} =
{\bm{\bar{J}}}_\mathcal{M}\cdot \bm{\tau}(t_0)$ is its footprint on the collection layer.
The generation of this footprint (see Fig. \ref{fig:ProjectionPlot}) results from the composition of two transformations, namely the three-dimensional
stretching as the falling surface is advected towards the collecting surface, $\bm{\tau}(t_z)=\bm{J}_\mathcal{M}\cdot \bm{\tau}(t_0)$, and its subsequent projection onto the horizontal along the direction of motion, $\bm{\bar\tau} =\mathcal{P} \bm{\tau}(t_z)$. The combination of these two processes gives the following relationship:
\begin{equation}
{\bm{\bar{J}}}_\mathcal{M} = \mathcal{P} \bm{J}_\mathcal{M} \ .
\end{equation}
This expression can be derived more formally by applying the
chain rule to Eq. (\ref{eq:rel}), as done explicitly in
\cite{Drotos2019} for the two-dimensional case.  Since the
singular values of $\mathcal{P}$ are $|\bv|/|v_z|$ and $1$,
standard inequalities for singular values of products of
matrices \cite{MatrixAnalysis} allow to show that
$\bar\Lambda_1\leq \Lambda_1 |\bv|/|v_z|$ and $\bar\Lambda_1
\bar\Lambda_2 \leq |\bv / v_z| \Lambda_1 \Lambda_2$. This last
inequality is however improved by the exact equality in Eq.
(\ref{eq:releigen}).

We now obtain expression (\ref{eq:SyP2}) for the projection
factor $P$ entering the density factor. We first note that
${\bm{\bar{\tau}_x}}  =\mathcal{P}\bm{\tau}_x (t_z)$ and
${\bm{\bar{\tau}_y}} =\mathcal{P}\bm{\tau}_y (t_z)$. Thus, we
can elaborate the expression for the density factor in Eq.
(\ref{eq:bartau_barlambda}) (we omit the time variable $t_z$ to
simplify the notation):
\begin{eqnarray}
F^{-1} &=& |{\bm{\bar{\tau}}_x}  \times {\bm{\bar{\tau}}_y}  | =
|\mathcal{P}\bm{\tau}_x  \times \mathcal{P}\bm{\tau}_y| \nonumber \\
&=& \left|\left(\bm{\hat z} \times \left(\bm{\tau}_x  \times \frac{\bv}{v_z}\right)\right)
\times \left(\bm{\hat z} \times \left(\bm{\tau}_y  \times \frac{\bv}{v_z}\right)\right)\right| \nonumber \\
&=&    \frac{1}{v_{z}^{2}}\left|(  \bm{\tau}_x (\bm{\hat z} \cdot \bv) - \bv (\bm{\hat z} \cdot \bm{\tau}_x )   \times (\bm{\tau}_y (\bm{\hat z} \cdot \bv) - \bv (\bm{\hat z} \cdot \bm{\tau}_y ) )\right|     \nonumber \\
&=&    \frac{1}{v_{z}^{2}}\left|(  \bm{\tau}_x v_{z} - \bv (\tau_x)_{z} ) \times (\bm{\tau}_y v_{z} - \bv (\tau_y)_{z} )\right|     \nonumber \\
&=&    \frac{1}{|v_{z}|}\left| v_{z}(\bm{\tau}_x  \times \bm{\tau}_y ) +  (\tau_x)_{z}(   \bm{\tau}_y  \times \bv ) -  (\tau_y)_{z}(\bm{\tau}_x  \times \bv )   \right|     \nonumber \\
&=& \left| \frac{ (\bm{\tau}_x  \times \bm{\tau}_y  ) \cdot \bv }{v_{z}} ~\bm{\hat z} \right|     \nonumber \\
&=&  \left|\frac{\hat\bn \cdot \bv}{v_z}\right| \left|\bm{\tau}_x  \times \bm{\tau}_y  \right|
\label{eq:ded}
\end{eqnarray}

Comparing with Eqs. (\ref{eq:eqF}) and (\ref{eq:SyP1}) we
identify the projection factor $P=|v_z/(\hat\bn \cdot \bv)|$,
thus demonstrating Eq. (\ref{eq:SyP2}).

\bibliography{FluidTransport.bib}

\end{document}